\documentclass[usegraphicx,useAMS,usenatbib,]{mn2e}
\usepackage{aas_macros}
\usepackage{graphicx}

\newcommand{\kms}{\mbox{${\rm km\,s}^{-1}$}}
\newcommand{\mss}{\mbox{${\rm m\,s}^{-2}$}}
\newcommand{\Mjup}{\mbox{${M}_{\rm Jup}$}}
\newcommand{\Rjup}{\mbox{${R}_{\rm Jup}$}}
\newcommand{\pjup}{\mbox{${\rho}_{\rm Jup}$}}
\newcommand{\Msun}{\mbox{${M}_{\sun}$}}
\newcommand{\Rsun}{\mbox{${R}_{\sun}$}}
\newcommand{\Msolar}{\mbox{${M}_{\sun}$}}

\newcommand{\psun}{\mbox{${\rho}_{\sun}$}}

\newcommand{\teff}{\mbox{$T_{\rm eff}$}}
\newcommand{\Teq}{\mbox{$T_{\rm eq}$}}
\newcommand{\logg}{\mbox{$\log g$}}
\newcommand{\vsini}{\mbox{$v \sin i$}}
\newcommand{\mictrb}{\mbox{$\xi_{\rm t}$}}
\newcommand{\mactrb}{\mbox{$v_{\rm mac}$}}

\newcommand{\halpha}{\mbox{$H_\alpha$}}
\newcommand{\hbeta}{\mbox{$H_\beta$}}

\newcommand{\mcc}[1]{\multicolumn{3}{c}{#1}}

\newcommand{\ermcc}[5]{\mcc{\ensuremath{{#1\,^{+#2}_{-#3}}\,^{+#4}_{-#5}}}}

\title[Spitzer lightcurves of WASP-18]
 { Spitzer 3.6{\boldmath $\mu$}m and 4.5{\boldmath $\mu$}m full-orbit
lightcurves of WASP-18}
\author[P.F.L. Maxted et~al.]
{P.F.L.~Maxted$^1$,
D.R.~Anderson$^1$,
A.~P.~Doyle$^1$,
M.~Gillon$^2$,
J.~Harrington$^3$,
\newauthor
N.~Iro$^1$,
E.~Jehin$^2$,
D.~Lafreni\`{e}re$^4$,
B.~Smalley$^1$,
J.~Southworth$^1$ \\
  $^1$Astrophysics Group,  Keele University, Keele, 
      Staffordshire ST5 5BG\\
 $^2$ Institut d'Astrophysique et de G\'{e}ophysique, Universit\'{e} de
Li\`{e}ge, All\'{e}e du 6 Ao\^{u}t 17, Bat. B5C, 4000 Li\`{e}ge, Belgium\\
 $^3$Planetary Sciences Group, Department of Physics, University of Central
Florida, Orlando, FL 32816-2385, USA\\
 $^4$D\'{e}partement de physique, Universit\'{e} de Montr\'{e}al, C.P. 6128 
Succ. Centre-Ville, Montr\'{e}al, QC, H3C 3J7, Canada \\
}

\date{Submitted 2012}

\pagerange{\pageref{firstpage}--\pageref{lastpage}} \pubyear{2008}

\def\LaTeX{L\kern-.36em\raise.3ex\hbox{a}\kern-.15em
    T\kern-.1667em\lower.7ex\hbox{E}\kern-.125emX}

\begin{document}
\label{firstpage}

\maketitle

\begin{abstract}
 We present new lightcurves of the massive hot Jupiter system WASP-18 obtained
with the Spitzer spacecraft covering the entire orbit at 3.6\,$\mu$m and
4.5\,$\mu$m. These lightcurves are used to measure the amplitude, shape and
phase of the thermal phase effect for WASP-18\,b. We find that our results for
the thermal phase effect are limited to an accuracy of about 0.01\,per\,cent
by systematic noise sources of unknown origin. At this level of accuracy we
find that the thermal phase effect has a peak-to-peak amplitude approximately
equal to the secondary eclipse depth, has a sinusoidal shape and that the
maximum brightness occurs at the same phase as mid-occultation to within about
5\,degrees at 3.6\,$\mu$m and to within about 10\,degrees at 4.5\,$\mu$m. The
shape and amplitude of the thermal phase curve imply very  low levels of heat
redistribution within the atmosphere of the planet. We also perform a separate
analysis to determine the system geometry by fitting a lightcurve model to the
data covering the occultation and the transit. The secondary eclipse depths we
measure at 3.6\,$\mu$m and 4.5\,$\mu$m are in good agreement with previous
measurements and imply a very low albedo for WASP-18\,b. The parameters of the
system (masses, radii, etc.) derived from our analysis  are in also  good
agreement with those from previous studies, but with improved precision. We
use new high-resolution imaging and published limits on the rate of change of
the mean radial velocity to  check for the presence of any faint companion
stars that may affect our results. We find that there is unlikely to be any
significant contribution to the flux at Spitzer wavelengths from a stellar
companion to WASP-18.  We find that there is no evidence for variations in the
times of eclipse from a linear ephemeris greater than about 100 seconds over 3
years.

\end{abstract}


\section{Introduction}
 Hot Jupiters are currently at the forefront of observational studies that can
provide meaningful tests for models of exoplanet atmospheres. The atmospheric
temperatures for a typical hot Jupiter orbiting a solar-type star with a
period of 3\,days can be up to 1500\,K. For transiting hot Jupiters this makes
it feasible to measure the planet-star flux ratio directly from the depth of
secondary eclipse in the lightcurve due to the occultation of the exoplanet by
the host star. Early results with the Spitzer Space Telescope confirmed the
existence of secondary eclipses in the lightcurves of HD\,209458
\citep{2005Natur.434..740D} and TrES-1 \citep{2005ApJ...626..523C} with the
expected depth $\sim 0.5$\,per\,cent at mid-infrared wavelengths. The secondary
eclipse depth has now been measured using Spitzer for more than 20 hot
Jupiters \citep{2011ApJ...729...54C}. Comparison of these observations with
atmospheric models has been used to reveal the diversity of hot Jupiter
atmospheres with regard to their composition \citep{2011Natur.469...64M}, the
presence or absence of a temperature inversion in their atmospheres,
\citep{2010ApJ...720.1569K}, and their albedos and heat recirculation
efficiencies \citep{2011ApJ...729...54C}.

 The secondary eclipse depth for a hot Jupiter at infrared wavelengths
measures the brightness temperature of the hemisphere facing the star -- the
``day-side'' -- integrated over the visible hemisphere.  This brightness
temperature, $T_{\rm day}$, will depend on the pattern of emission over the
day side, the spectral energy distribution (SED) of this emission, the Bond
Albedo, $A_b$, and the efficiency with which heat is redistributed to the
night-side of the planet. Observations at several wavelengths, particularly
near-infrared observations near the peak of the day-side SED, reduce the
extent to which we must rely on models to account for the conversion from
brightness temperature to effective temperature when interpreting these
observations \citep{2011Natur.469...64M}. In general, it is not possible to
disentangle the degeneracy between Bond albedo and heat recirculation
efficiencies from secondary eclipse observations alone. One exception is the
case of HD\,189733, in which very high quality Spitzer observations of the
secondary eclipse reveal asymmetries that can be inverted to produce a map of
the brightness temperature on the day side of HD\,189733\,b
\citep{2012ApJ...747L..20M,  2012arXiv1202.3829D}. Apart from this exceptional
case, it is currently only possible to obtain information on the
redistribution of heat in the planet's atmosphere by observing the thermal
phase effect -- the variation in infrared brightness of the system as a
function of orbital phase.

 There are several different ways to parametrize the redistribution of heat
from the day-side to the night-side of a planet \citep{2010ApJ...722..871S}.
Here we use the parameter $P_n$, the fraction of the incident energy that is
transported to the night side of the planet. Plausible values of this
parameter vary from $0$ up to $0.5$. A value of $P_n=0$ would imply a
night-side brightness temperature $T_{\rm night} \ll T_{\rm day}$, whereas
$P_n=0.5$ implies $T_{\rm night} \approx  T_{\rm day}$.\footnote{The
brightness temperatures are not necessarily equal since the SED from the
day-side and night-side may be different.} If $P_n > 0$ then this suggests
that winds at some level in the atmosphere move heat around the planet.  In
practise, very high efficiencies  for strongly irradiated planets are
unlikely because the winds that transport heat to the night-side will
dissipate some of their energy through turbulence or shocks
\citep{2009ApJ...693.1645G}. Nevertheless, some redistribution of energy from
the day-side to the night-side is likely, and may lead to significant offsets
between the sub-stellar point and the hottest regions of the atmosphere
\citep{2005ApJ...629L..45C}. This will be observed in the thermal phase effect
as an offset in the phase of maximum brightness from opposition.
  
\citet{2007MNRAS.379..641C} used 8 separate observations  with Spitzer  spread
throughout the orbit of the three hot Jupiter systems HD\,209458, HD\,179949
and 51~Peg to measure their thermal phase effect. They were able to place
useful upper limits on the phase variation in 51~Peg and HD\,209458 and to
detect a variation with a peak-to-trough amplitude of 0.14\,per\,cent in
HD\,179949. HD\,179949 is a non-transiting hot Jupiter, so the inclination of
the orbit and the radius of the planet are unknown, but even allowing for this
uncertainty, the observed amplitude of the phase variation provided an upper
limit of $P_n < 0.21$ and shows that the hottest point is near the sub-stellar
point. HD\,209458 is a transiting system, so the inclination of the orbit and
the radius of the exoplanet are known. This allowed
\citeauthor{2007MNRAS.379..641C} to translate the upper limit on the amplitude
of the thermal phase into a lower limit $P_n > 0.32$, and thus establish that
apparently different hot Jupiters are likely to have  a variety of $P_n$
values. 

 \citet{2006Sci...314..623H} detected the phase variation  of the planet
$\upsilon$~And\,b using the MIPS instrument on Spitzer at 24$\mu$m. With
additional data and an improved understanding of the systematic noise sources
in MIPS, they were able to refine their estimate of the amplitude of the phase
variation and show that  there is a large ($\sim80^{\circ}$) phase offset
between the time of maximum brightness and opposition
\citep{2010ApJ...723.1436C}. The inclination and radius of  $\upsilon$~And\,b
are not known accurately because it is a non-transiting exoplanet. This
can result in a large uncertainty in the value of $P_n$ inferred from the 
amplitude of the phase curve \citep{2008ApJ...678.1436B}. However, the large
phase offset observed in $\upsilon$~And\,b implies a large $P_{\rm n}$
despite the large amplitude of the phase variation. 

 The thermal phase curve of HD189733\,b has been observed using Spitzer at
3.6$\mu$m and 4.5$\mu$m \citep{2012ApJ...754...22K}, at 8$\mu$m
\citep{2007Natur.447..183K} and at 24$\mu$m \citep{2009ApJ...690..822K}. There
are also multiple observations of the transits and eclipses at 8$\mu$m for
this system \citep{2010ApJ...721.1861A}. The combined analysis of these
results by \citet{2012ApJ...754...22K} shows that heat recirculation from the
day-side to the night-side  is efficient for this relatively cool hot Jupiter
($T_{\rm day} \approx 1200$\,K) and that this recirculation leads to a peak in
the thermal phase effect that occurs $\sim 25$\,degrees before opposition.

 \citet{2012ApJ...747...82C} obtained Warm Spitzer photometry covering the
complete 26\,hour orbit of the very hot Jupiter WASP-12\,b at 3.6\,$\mu$m and
4.5\,$\mu$m. They found that their interpretation of the lightcurves depends
on the assumptions made about the nature of the systematic noise in the
lightcurves and that red noise is the dominant source of uncertainty in their
analysis. Nevertheless, they were able to show that the thermal phase
variation in WASP-12\,b is large, indicative of poor day-to-night heat
redistribution ($P_n \la 0.1$). The small offset they observe between the
phase of maximum brightness and secondary eclipse ($16\pm4$\,degrees) in the
4.5\,$\mu$m data is consistent with this interpretation. The phase offset at
3.6\,$\mu$m could not be determined unambiguously from their data.

 Although thermal phase curves are only currently available for a few systems,
there does appear to be a pattern of weak recirculation for the hottest
planets. \cite{2011ApJ...729...54C} have looked for trends in the value of
$T_{\rm day}/T_{\rm sub}$, where $T_{\rm sub}$ is the equilibrium temperature
of the sub-stellar point, in a sample of 24 transiting exoplanets with
secondary eclipse measurements. This quantity will depend on both the albedo
of the planet and the recirculation efficiency, but a large value can only be
obtain if both the albedo and the recirculation efficiency are low.
\citeauthor{2011ApJ...729...54C} found that this is the case for all of the 6
hottest planets ($T_{\rm day} \ge 2400$\,K) in their sample and point out that
this is, in general terms, the expected behviour given that the radiative
timescale scales as $T^{-3}$ whereas the advective timescale (which they
assume to be of-order the local sound speed) scales  as $T^{-0.5}$. This
simple scaling argument does not explain the apparent transition in behaviour
at $T_{\rm day} \approx 2400$\,K but \citet{2012ApJ...751...59P} do observe
a transition at about this temperature in their suite of three-dimensional
circulation models for hot Jupiters. This is mainly due to the change in the 
ratio of the radiative and advective timescales, with the presence of an
atmospheric inversion playing a lesser role in determining the recirculation
efficiency. This transition may also be related to the onset of ionisation of
alkali metals in  the planet's atmosphere, leading to severe magnetic drag
\citep{2010ApJ...724..313P}. There is an ongoing debate as to whether the
resulting Ohmic dissipation  can transport energy into the interior of the
planet and so explain the very large radii observed for some hot Jupiters
\citep{2011ApJ...729L...7L, 2012ApJ...750...96R, 2012ApJ...751...59P,
2012ApJ...757...47H}. 

 Here we present Warm Spitzer photometry covering the complete orbit of the
very hot Jupiter WASP-18\,b. This exoplanet is unusual for its combination of
short orbital period (0.945\,d) and high mass (10\,M$_{\rm Jup}$), which
results in strong tidal interactions between the planet and the star  
\citep{2009Natur.460.1098H}. \citet{2009ApJ...707..167S} derived accurate
masses and radii for the star and planet in the WASP-18 system based on high
quality optical photometry of the transit and the spectroscopic orbit from
\citeauthor{2009Natur.460.1098H}. \citet{2011ApJ...742...35N} used  Spitzer
photometry in all four IRAC bands covering the secondary eclipse of WASP-18 to
measure the brightness temperature of the day side from 3.6\,$\mu$m to
8.0\,$\mu$m. The high brightness temperatures derived ($\sim 3200$\,K) imply
that WASP-18\,b has near-zero albedo and almost no redistribution of energy
from the day side to the night side of the planet.

 Our primary aim is to use our Warm Spitzer photometry at 3.6\,$\mu$m and
4.5\,$\mu$m  to measure the amplitude, phase and shape of the thermal phase
effect. We also use these data to re-measure the secondary  eclipse depths at
3.6\,$\mu$m and 4.5\,$\mu$m for comparison with the results of
\citet{2011ApJ...742...35N}. We measure the times of eclipse and transit from
our data and from new optical photometry of several transits and use these
together with published times of mid-eclipse to re-measure the eclipse
ephemeris and to look for possible variations in the period. We consider the
likelihood that WASP-18 has a stellar companion based on published radial
velocity data and new high-resolution imaging at H-band and K-band. The
contamination of the lightcurve for a hot Jupiter system by a companion star
has the potential to bias the results obtained if not properly accounted for.
Companions stars may also play a role in the formation and evolution of some
hot Jupiter systems \citep{2007ApJ...669.1298F, 2007MNRAS.382.1768M}. Our
analysis also provides an accurate characterisation of the primary eclipse
(transit) which can be used in combination with other data to re-measure the
mass and radius of the star and planet.

\section{Observations}
\subsection{Spitzer photometry}

 We were awarded {\it Spitzer} General Observer time during
Cycle~6\footnote{PI: P. Maxted, program ID 60185} to observe two complete
orbits of WASP-18 with IRAC \citep{1998SPIE.3354.1024F}, one orbit with each
of the two IRAC channels operating  during the warm mission.  Observations
with channel 1 (3.6$\mu$m) were obtained on 2010 January 23, observations with
channel 2 (4.5$\mu$m) were obtained on 2010 August 23.  On both dates,
243\,200 images with an exposure time of 0.36s were obtained in sub-array
mode. The total duration of each sequence of observations is 29 hours.  In
addition, a sequence of 64 sub-array images also with an exposure time of
0.36s were obtained immediately after the observations of WASP-18  at a
slightly offset position. These were used to check for hot pixels or other
image artifacts on the detector. We used Basic Calibrated Data (BCD) processed
with version S18.18 of the Spitzer IRAC pipeline for our analysis.

\subsection{AO imaging\label{ao}}
 We obtained adaptive optics high resolution $H$- and $K$-band images  of
WASP-18 using the NICI instrument at Gemini South. The observations  were
obtained on the night of 2010 December 27 under good seeing
(0.5\,--\,0.6\,arcsec). The instrument was configured with the CLEAR focal
plane mask and the H~50/50 beam-splitter, and we used the narrow band  filters
$K_{\rm cont}$ (2.2718~$\mu$m) and FeII (1.644~$\mu$m) in the  red and blue
channels, respectively. We observed the target at five  dither positions
corresponding to the center and corners of a square  of side 6\,arcsec. At
each dither position we obtained three images  consisting of the co-addition
of 3 exposures of 1.5~s. The full-width at half-maximum (FWHM) of the point
spread function (PSF) in these images was 0.065\,arcsec in the FeII filter and
0.073\,arcsec in the $K_{\rm cont}$ filter.

 Data reduction consisted of subtracting a sky image, dividing by the  flat
field, and fixing bad pixels by interpolating from neighbouring  pixels. The
sky image was created from the median combination of the images at the
different  dither positions after masking out the star signal in each image.
The  reduced images were registered to a common position and field
orientation and then combined using the median value of each pixel.  No other
point source was  detected in the resulting images. The sensitivity of our AO
imaging to  detect faint companions was determined by first computing the
median  absolute deviation of the pixel values within annuli of various radii
and width equal to one PSF FWHM. The resulting contrast curve was then
properly scaled, and verified to be adequate, by adding and recovering  (by
visual inspection) fake companions in the images at various  separation and
with various contrasts. In doing this last exercise we  used both the $K$- and
$H$-band images to differentiate speckles from  true companions, which display
a different chromatic behaviour. Using  this approach, we estimate the
detection limits in difference of  magnitudes to be 4.0~mag at
$\ge$0.2\arcsec, 5.4~mag at $\ge$0.4\arcsec \ and 6.0~mag at $\ge$0.5\arcsec.
Finally, we note the presence of a  faint ghost in the image at 13~pixel
(0.23\arcsec) separation and  contrast of $\sim$4.1~mag in the $K_{\rm cont}$
filter and $\sim $5.9~mag in the FeII filter.

\subsection{TRAPPIST photometry}
Five transits of WASP-18\,b were observed with the 60\,cm robotic telescope
TRAPPIST\footnote{http://www.ati.ulg.ac.be/TRAPPIST}
\citep{2011Msngr.145....2J, 2012A&A...542A...4G}
 located at ESO La Silla Observatory (Chile). TRAPPIST is
equipped with a thermo-electrically-cooled 2k\,$\times$\,2k CCD camera with a
field of view of $22^{\prime}\,\times\,22^{\prime}$ (pixel scale =
0.65\,arcsec\,pixel$^{-1}$). The first four
transits were observed
in an Astrodon `$I+z$' filter that has a transmittance $> 90$\,per\,cent from
750\,nm to beyond 1100\,nm, the red end of the effective bandpass being defined
by the spectral response of the CCD. The last transit was observed in the
Sloan $z$' filter. For all transits, the telescope was slightly defocused to
minimize pixel-to-pixel effects and to optimize the observational efficiency.
For each run, the stellar images were kept on the same pixels, thanks to
`software guiding' system deriving regularly astrometric solutions on the
science images and sending pointing corrections to the mount if needed.
Table~\ref{wasp18-phot} gives the logs of these TRAPPIST observations. 

 After a standard pre-reduction (bias, dark, flatfield correction), the
stellar fluxes were extracted from the images using the {\sc
iraf/daophot}\footnote{{\sc iraf} is distributed by the National Optical
Astronomy Observatory, which is operated by the Association of Universities
for Research in Astronomy, Inc., under cooperative agreement with the National
Science Foundation.} aperture photometry software \citep{1987PASP...99..191S}.
For each transit, several sets of reduction parameters were tested, and we
kept the one giving the most precise photometry for the stars of similar
brightness as WASP-18. After a careful selection of reference stars,
differential photometry was  obtained.

\begin{table}
\caption{Details of the transit lightcurves obtained with TRAPPIST for
WASP-18. For each lightcurve we list the observation date, the filter
used, the number of measurements, and the exposure time. }
\begin{center}
\begin{tabular}{lcrr}
\hline \noalign {\smallskip} 
Date  &   \multicolumn{1}{l}{Filter} & \multicolumn{1}{l}{$N_p$}  & 
\multicolumn{1}{l}{T$_{\rm exp}$ [s]}  \\
\hline \noalign {\smallskip} 
2010 Sep 30 &  {\it I+z} & 712 & 12  \\ \noalign {\smallskip}
2010 Oct 02 &  {\it I+z} &  977  & 8  \\ \noalign {\smallskip}
2010 Dec 23 &  {\it I+z} & 688 & 6  \\ \noalign {\smallskip}
2011 Jan 08 &   {\it I+z} & 648 & 6  \\ \noalign {\smallskip}
2011 Nov 10 &  {\it z'} & 624 & 10  \\ \noalign {\smallskip}
\hline
\end{tabular}
\end{center}
\label{wasp18-phot}
\end{table}
 
\section{Spitzer Data analysis}

\subsection{Conversion to flux and noise model} 
 We converted the BCD images from units of MJy/steradian to mJy using the
values for the pixel size at the centre of the subarray provided in the image
headers ($1.225^{\prime\prime}\times1.236^{\prime\prime}$ for channel 1, 
$1.205^{\prime\prime}\times1.228^{\prime\prime}$ for channel 2). We used the raw
images together with the values of the gain and readout noise for each channel
to calculate the noise level in each pixel assuming Poisson counting
statistics. 

\subsection{Image times}
 BCD data in sub-array mode are delivered as FITS files containing a data cube
of 64 images of $32\times32$ pixels per file.  We used the FITS header keyword
BMJD\_OBS to assign a Barycentric UTC modified Julian date (BMJD) to the start
of the exposure for the first image in the data cube.  The BMJD of the
mid-exposure time for each image in the data cube was  then calculated using
the values for the start and end times of the integration from the FITS header
(AINTBEG and ATIMEEND) to calculate the time taken to obtain the 64 images
and assuming that these images were uniformly spaced in time.  

 Long observing sequences such as the ones we have used for WASP-18 cannot 
be executed using standard observing modes so multiple instrument engineering
requests (IERs) are used to obtain the data. The images obtained in the second
of the two IERs used for our WASP-18 observations do not have the coordinates
of the target in the FITS header and so the light-time correction to the solar
system barycentre are incorrect for these images. We calculated the light-time
correction for the images obtained before the interruption from the difference
in the keyword values BMJD\_OBS $-$ MJD\_OBS. We then use a linear
extrapolation of this light-time correction as a function of MJD\_OBS  to
calculate the BMJD of the images obtained after the interruption based on
their MJD\_OBS values. The uncertainty in the exposure time introduced by this
procedure is negligible. 

\subsection{Outlier rejection}
 We compared each image to the other 63 images in the same data cube in order
to identify discrepant data points in the images. We are particularly
concerned here with identifying discrepant pixel values that may affect the
photometry of the target. As the target moves during the sequence of 64
images, we use a robust linear fit (least absolute deviation) to the 64 pixel
values from each file for each pixel to predict the expected value for each
pixel value in each image. We then flag the pixels in each image that deviate
from their expected value by more than 5 times their standard error.  We find
that the number of pixels flagged using this method is much larger than
expected given the known incidence of cosmic ray hits on the IRAC detectors.
This discrepancy is due to a few pixels well away from the target position
that are noisier than predicted by our noise model. As these pixels have no
effect on our photometry and a negligible effect on the estimate of the
background level we ignore this discrepancy. We also flagged any pixels in our
images that are flagged as bad pixels in the ``Imask'' file provided for each
BCD file by the Spitzer IRAC pipeline. 

\subsection{Sky background estimate} 
 We use the mean of the  image pixel values excluding those within 10
pixels of the target position to estimate the background value in each image.
Values more than 4 standard deviations from the mean and flagged pixels were
ignored in the calculation. We used a Gaussian fit to a histogram of these
pixel values to estimate the standard deviation of the  background pixel
values, $\sigma_{\rm bg}$. The number of points used to estimate the
background was $\approx 700$. Typical values of $\sigma_{\rm bg}$  are
0.0033\,mJy/pixel for channel 1 and 0.0025\,mJy/pixel for channel 2.

\subsection{Aperture photometry} 

 We tried three different methods to measure the location of the star on the
detector, the {\sc daophot} {\sc cntrd} and {\sc gcntrd} algorithms and a
least-squares fit of a bivariate Gaussian distribution to an 11$\times$11
sub-image centred on the nominal star position. We refer to this latter
algorithm as the {\sc gauss2d} method. We used a fixed value for the
full-width at half-maximum of FWHM$=1.25$ pixels for both axes of the Gaussian
profile in the {\sc gauss2d} method based on the results of fitting the images
with the FWHM as a free parameter. The {\sc cntrd} algorithm determines the
position where the derivatives of the image values go to zero. The {\sc
gcntrd} algorithm fits a Gaussian profile to the marginal $x$ and $y$
distributions of the image values. We set the input parameters to {\sc cntrd}
and {\sc gcntrd} to run on a sub-image of $5\times5$ pixels around the target
position. We compare the performance of these different algorithms below.   

 We used the IDL Astronomy Users library\footnote{\it
http://idlastro.gsfc.nasa.gov/} implementation of the {\sc daophot} {\sc aper}
procedure \citep{1987PASP...99..191S} to perform synthetic aperture photometry
on our images. We used the 2006 November version of this procedure which
allowed us to use the option to use an exact calculation of the intersection
between a circular aperture and square pixel for correct weighting of pixels
at the edge of the aperture. We used 13 aperture radii uniformly distributed
from 1.5  pixels to 4.5 pixels. The results we obtained for larger aperture
radii were not useful because the lightcurves have much lower signal-to-noise
due to the additional background noise. Fluxes measured from images containing
any flagged pixels in the aperture  were rejected from further analysis,
although much less than 1\,per\,cent of images were affected in this way. The
median number of rejected pixels per image is 2. 

\subsection{Image persistence}
\begin{figure} 
\includegraphics[width=0.45\textwidth]{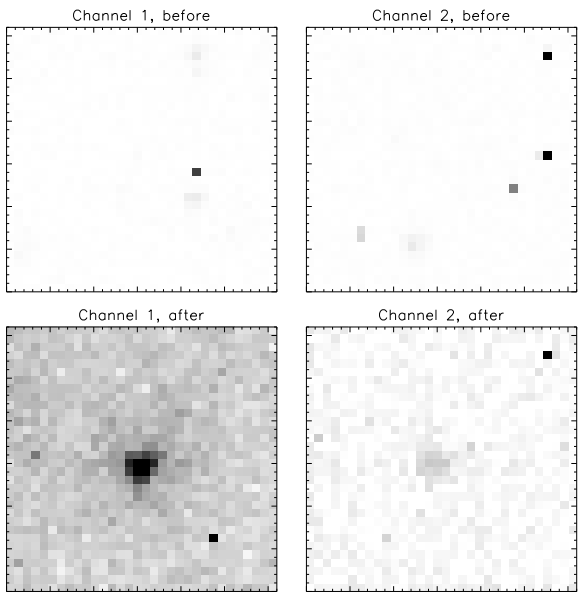} 
\caption{Images obtained before and after our WASP-18 observations. All images
are linearly scaled (inverse grey scale) between 0 and 10 MJy/sr. The
``before'' images are 30\,s exposures in the region of IC\,2560. The ``after''
images are the median of the 64 offset sky images with an exposure time of
30\,s each.
\label{afterimage}}  
\end{figure} 

 Our IRAC images are affected by image persistence, particularly the channel 1
images. This can be seen in the offset sky images obtained immediately after
our WASP-18 observations (Fig.~\ref{afterimage}). The resulting image
artifacts are more diffuse than the stellar images and look more like the
logarithm of the point spread function, as described in the IRAC instrument
handbook (Version 2.0.1, p. 116) . The image artifact in the channel 1 offset
sky images has up to about 0.6 mJy per pixel in the channel 1 image and a
total of about 6\,mJy within an aperture with a radius of 5 pixels. For
comparison, a typical channel 1 image of WASP-18 has a peak flux of
50\,--\,70\,mJy/pixel and a total flux of $163\pm3$\,mJy  within an aperture
of the same radius, so the image artifact affects the photometry of WASP-18 by
a few per\,cent. For the channel 2 data the corresponding figures are a total
of about 1\,mJy/pixel in the artifact compared to 42\,mJy/pixel in the peak
and a total flux of $103\pm1$\,mJy in the images of WASP-18 so for this
channel image persistence affects the photometry by about 1\,per~cent.
 
 The IRAC instrument handbook describes the behaviour of image persistence
artifacts in the IRAC arrays  during the warm mission. In channel 1 the
artifacts decay exponentially with a timescale of about 4.5\,hours. Channel 2
residuals start out as positive, but then become negative with a decay
timescale of a few minutes. 

 For channel 1 data we make the assumption that the image artifact will be
approximately constant after some time during the exposure sequence comparable
to the decay timescale. To correct for the effect of image persistence we
create a ``master offset image'' from  the median of the 64 offset sky images,
subtract the background value from this image and then subtract the result
from the images of WASP-18. This correction will be inaccurate for some
fraction of the data at the start of the observing sequence while the image
persistence builds-up. We discuss this point further below. 

 For the channel 2 data it is not clear how the image artifact affects the
photometry of WASP-18. The interval between the end of the observing sequence
for WASP-18 and the start of the offset sky image is 49\,s, which is
comparable to the decay timescale for the artifact. The exact form and
timescale for the decay of the artifact is not known so it is not even
possible to make a precise estimate of contribution of the image artifact to
the measured flux for WASP-18. However, it is likely that the image artifact
contributes less than 2\,per\,cent given the decay timescale for this feature
is a few minutes. We did attempt to measure the decay timescale from the data
taken subsequent to our own observations, but the artifact was not detectable
in those images. For the channel 2 data we treat the contribution of the image
persistence artifact as an additional source of uncertainty in our analysis.

\subsection{Initial assessment of the data}
\begin{figure*} 
\includegraphics[width=0.95\textwidth]{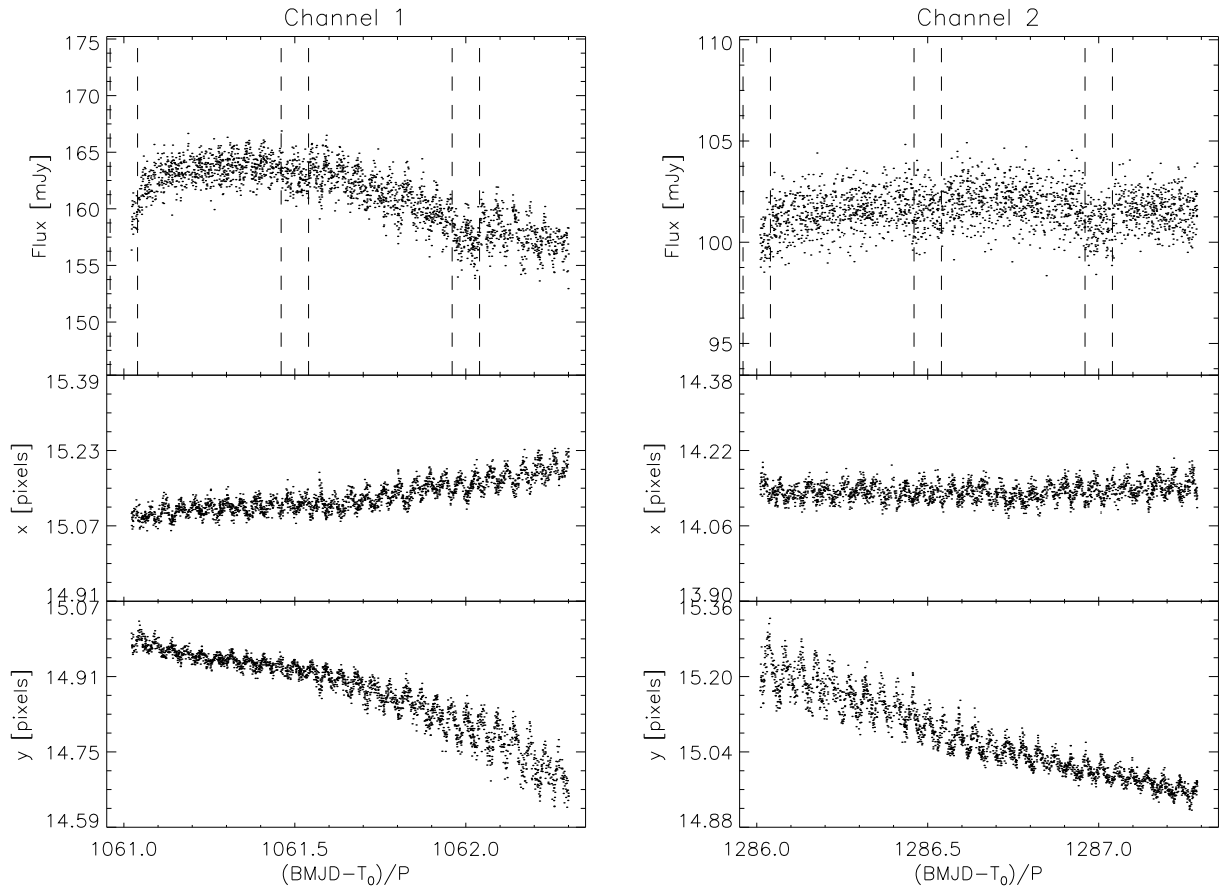} 
\caption{The flux of WASP-18 measured in IRAC channel 1 (left panel) and
channel 2 (right panel) measured with a circular aperture of radius of 4.5
pixels.. The position of the star on the array measured using the {\sc
gauss2d} method is shown below each panel. For
clarity, we have only plotted a random selection of 1 per\,cent of the data
here.   Dashed lines indicate the start and end times of the transit and
secondary eclipse assuming a duration of 0.08\,d for each and assuming that
the secondary eclipse occurs at phase 0.5.
\label{raw}}  
\end{figure*}

 The flux of WASP-18 measured with an aperture radius of 4.5 pixels is shown
for both channels in Fig.~\ref{raw}.   Also shown in this figure are the
positions of the star on the array calculated using the {\sc gauss2d} method.
The coordinates $x$ and $y$ are measured relative to the centre of a corner
pixel in the sub-array. The form of the variations in the $x$,$y$ positions
measured using the {\sc gcntrd} and {\sc cntrd} algorithms are similar, but
the amplitudes of the variations are less and there is an offset between these
values and the results of the {\sc gauss2d} method. For example, the $y$
positions measured for the channel 1 data using the {\sc gcntrd} method have
median value of 14.95 with 98\,per\,cent of the data in the range $y =
14.78$\,--\,$14.95$, cf. a median value of 14.88 and  range $y =
14.65$\,--\,$14.99$ for the {\sc gauss2d} method.

 The feature that stands out from Fig.~\ref{raw} is the well-known correlation
between the measured flux and the position of the star on the
detector, particularly in channel 1. This position-dependent sensitivity
variation (PDSV) makes it difficult to see the transit and secondary eclipse
in these ``raw'' aperture flux measurements. The channel 2 data appear to be
less affected by PDSV, so the  transit and secondary eclipse can be seen in
the raw flux measurements.  PDSV is a combination of the ``pixel phase
effect'' described in the IRAC instrument handbook (Version 2.0.1, p. 45) and
pixelation noise. The pixel phase effect is a variation in the sensitivity of
each detector pixel that depends on the distance of the stellar image from the
centre of the pixel. The motion in the $x$ and $y$ directions for our channel
2 data result in a smaller variation in the distance of the star from the
centre of the pixel compared to the channel 1 data, which may partly explain
why the data quality is better in this channel \citep{2011MNRAS.416.2108A}.
Pixelation noise affects synthetic aperture photometry with small aperture
radii because for a pixel at the edge of the aperture, the fraction of the
flux falling within that pixel is not the same as the fraction of the aperture
within the pixel.

\begin{figure*} 
\includegraphics[width=0.95\textwidth]{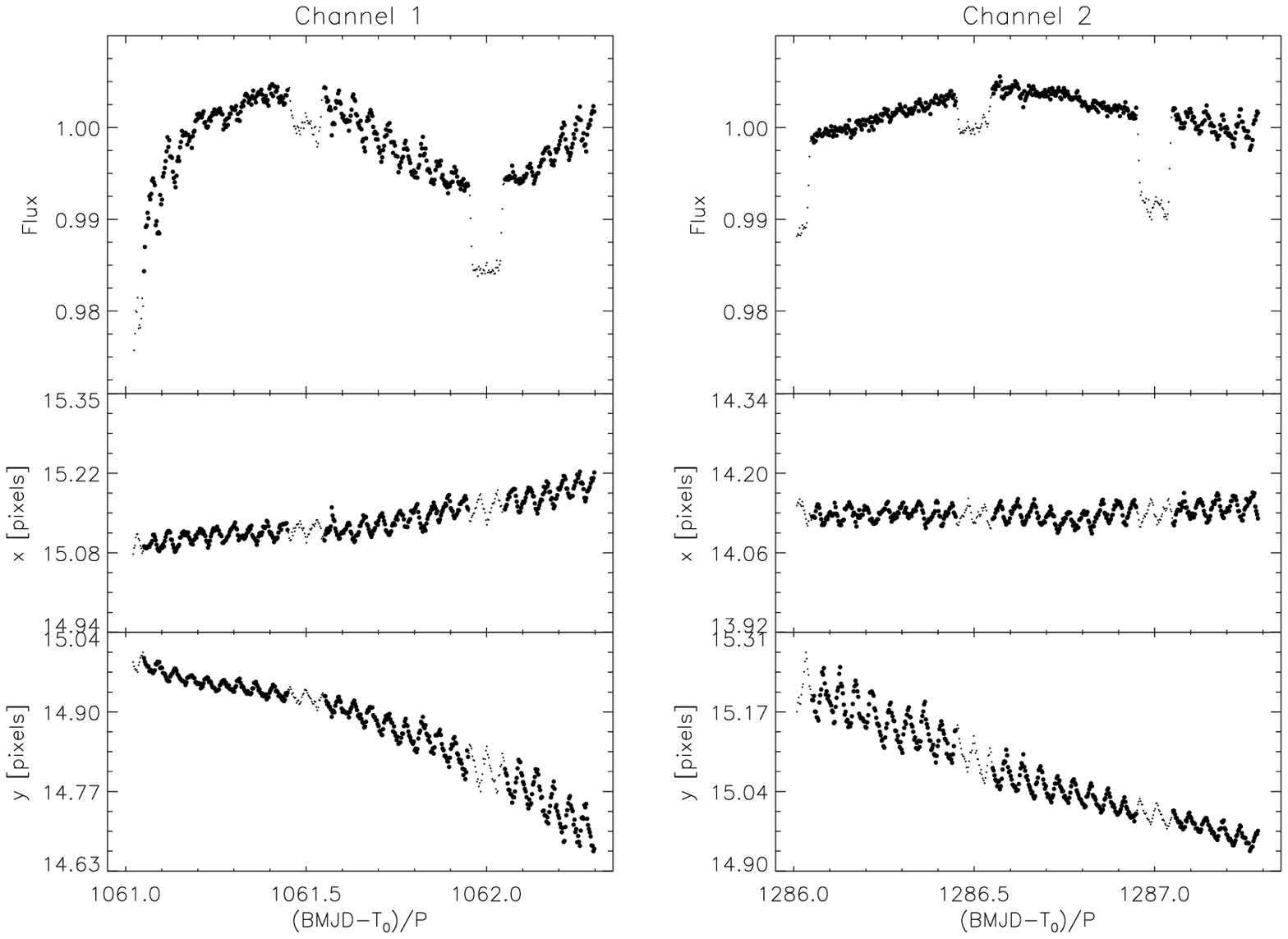} 
\caption{The flux of WASP-18 measured in IRAC channel 1 (left panel) and
channel 2 (right panel) after a linear correction for position dependent
sensitivity variations and a simple sinusoidal model for the phase variation.
The zero-point of the flux scale is set from the mean flux during
secondary eclipse. Observations obtained during the transit and secondary
eclipse (small points) were excluded from the calculation of the coefficients
for the decorrelation. The data have been binned into 0.0025\,d bins for
display purposes only. \label{decor11110000}}  \end{figure*} 

\subsection{Analysis of the thermal phase effect}
\subsubsection{Correction for PDSV}
 The usual method developed to correct for PDSV in IRAC
data for observations of the secondary eclipses of hot Jupiters is
to include parameters in the least-squares fit of an eclipse  model to the
data to represent the PDSV. This is usually a simple linear or quadratic
relation between sensitivity and each of the coordinates $x$ and $y$ (e.g.,
\citealt{2011ApJ...727...23B}; \citealt{2011MNRAS.416.2108A}).
\cite{2010PASP..122.1341B} have developed an alternative method to correct for
PDSV in their Warm Spitzer 4.5$\mu$m observations of GJ\,436. They created a
{\it pixel sensitivity map} from the data themselves. This approach was
straightforward in the case of GJ\,436 because the lightcurve of the target is
expected to be constant apart from the possible presence of transits affecting
a small fraction of the data.  The pixel sensitivity map generated by
\citeauthor{2010PASP..122.1341B} for the IRAC channel 2 shows complex
structure that they describe as ``corrugation \dots\  low-level
sinusoidal-like variations with a separation of approximately 5/100ths of a
pixel between peaks''. A similar concept based on bi-linear interpolation
rather than a smoothed look-up table has been developed by
\citet{2012ApJ...754..136S} and applied to Spitzer photometry of HD\,149026\,b.

For our WASP-18 data we are interested in characterising the amplitude and
shape of the phase variation as well as measuring the shape and depths of the
transit and secondary eclipse. The phase variation has a period comparable to
the length of the observing sequence so it is important to understand any
correlations between the correction for the PDSV and the parameters of the
lightcurve model. The phase variation of WASP-18 can be modelled approximately
as a sinusoidal variation in flux with P=0.94\,d. For example, in the worst
case scenario, if the position of the target on the detector also varied
approximately sinusoidally with approximately the same period,  then it would
become impossible to determine whether any variation in the measured flux with
$P\approx0.94$ is due to the flux variation of WASP-18 or due to the PDSV.

Understanding the correlations between the lightcurve model parameters and
the correction for PDSV is problematic in the case of our WASP-18 data because
of the large number of parameters required to model the complete lightcurve
plus the large number of parameters that may be required to characterise the
complex structure (``corrugation'') in the PDSV. One method we experimented
with was to  use the pixel sensitivity map method of \cite{2010PASP..122.1341B}
applied to the residuals to a least-squares fit of a lightcurve model. This
approach can be applied iteratively until the solution and pixel sensitivity
map converge. The problem with this approach is that it becomes difficult to
identify correlations between the lightcurve model parameters and the pixel
sensitivity map. We avoided this problem by excluding the data during the
transit and occultation from our analysis of the thermal phase effect. The
main advantage of this approach is that fitting a model to the data between
the eclipses can be reduced to a  linear least-squares problem. This makes is
straightforward to   find the best solution of the problem and to investigate
the correlations between the free parameters of the model.

\subsubsection{Model for PDSV and the thermal phase effect}
Our model  for the measured magnitude of the system between the eclipse and
transit is 
\begin{equation}
\begin{array}{lll}
m_i & = & c_{0,0} 
+   \sum_{j=1}^{N_{\cos}} a_j \cos(j\phi_i) 
+   \sum_{k=1}^{N_{\sin}} b_k \sin(k\phi_i)  \\
\noalign{\smallskip}
&& +  \sum_{\iota=1}^{N_{x}} c_{\iota,0} p_{\iota} (x^{\prime}_i)
+   \sum_{\kappa=1}^{N_{y}} c_{0,\kappa} p_{\kappa} (y^{\prime}_i)  \\
\noalign{\smallskip}
&& +  \sum_{\lambda=1}^{N_{xy}}\sum_{\mu=1}^{N_{xy}} 
   c_{\lambda,\mu} p_{\lambda} (x^{\prime}_i) p_{\mu} (y^{\prime}_i), \\
\end{array}
\end{equation}
where $m_i$ is the magnitude of WASP-18 at time $t_i$; $\phi_i =
2\pi(t_i-T_0)/P$ is the orbital phase relative to the time of mid-transit,
$T_0$;  $p_n$ is a Legendre polynomial of order $n$; $x^{\prime}_i =
(x_i-\bar{x})/(x_{\rm max} - x_{\rm min})$ and similarly for
$y^{\prime}_i$ ($x_{\rm min}$ is the minimum value of $x_i$, etc.).  We use
the values $T_0={\rm BMJD~} 54220.98163$ for the time of mid-transit and
$P=0.94145299$\,d for the orbital period from \cite{2009Natur.460.1098H}
throughout this paper unless otherwise stated. By using Legendre polynomials
and normalized coordinates $(x^{\prime}_i,y^{\prime}_i)$ we find that we can
use singular value decomposition to find  solutions of this least-squares
problem for Legendre polynomials up to at least 12$^{\rm th}$ order. This is
sufficient to model the corrugations with a scale of 0.05 pixels seen by
\citeauthor{2010PASP..122.1341B} if they are present in our data. 

\subsubsection{Linear decorrelation against position}
In Fig.~\ref{decor11110000} we show the result of using the simplest
reasonable model for our data, in which the magnitude of the phase variation
varies sinusoidally and the PDSV is linear in $x$ and $y$, i.e., 
\[ m_i  =  c_{0,0} + a_1\cos(\phi_i) + b_1\sin(\phi_i) + c_{1,0}p_1(x^{\prime}_i) + 
c_{0,1}p_1(y^{\prime}_i).\]
Note that our calculations are done using magnitudes, but we plot the results
as fluxes and quote parameter values in unit of  per\,cent. The least-squares
fit of this model to the unbinned aperture photometry outside of eclipse and
transit is used to determine the coefficients of the model for the PDSV.  We
then apply this correction to all the data. The results in
Fig.~\ref{decor11110000} are for an aperture radius of 2.5 pixels for both
apertures and target positions measured using the {\sc gauss2d} method. This
is the combination of aperture radius and positions that gave the lowest RMS
residual for the data between the transit and eclipse. Results for other
apertures and for {\sc cntrd} and {\sc gcntrd} methods are similar. It is
clear that a linear correction is insufficient to fully remove the effect of
the PDSV, but this simple model does show clearly some features of our data.
Firstly, we see that the eclipse and transit are clearly visible in both
channels. Two transits are visible in the channel 2 data but the first transit
is not seen clearly in the channel 1 data because there is a large ``ramp''
affecting the first few hours of the data. There is a cosine-like variation in
flux observed in both channels with the maximum flux occurring near phase 0.5
(secondary eclipse). Part of this signal is the phase variation we wish to
measure. However, there must also be some instrumental component or other
systematic effect that contributes to this variation because the phase
variation due to the planet cannot have an amplitude larger than the secondary
eclipse depth.

The obvious suspect for the systematic noise source in the channel 1 data is
the image artifact shown in Fig.~\ref{afterimage}. The ``ramp'' is the right
size ($\approx 3$\,per\,cent) and builds up over the same sort of timescale as
the known decay timescale of this artifact. Our interpretation of this
lightcurve is that the image artifact builds up over the first 6\,--\,8 hours
before reaching an approximate equilibrium between the arrival of new photons
from WASP-18 and its own decay timescale.

We have tried several methods to account for this ramp-like feature in the
data but none of these methods is any better than the more pragmatic approach
of simply excluding the first 6\,--\,8 hours of the channel 1 data. Without
going into the details of these various methods, we can state here that we
almost always found that the amplitude of the phase variation measures in
channel 1 was similar to  the depth of the secondary eclipse and often was
slightly larger. It is possible to create models for the systematic noise in
the channel 1 lightcurves that achieve more physically realistic (lower)
values for the amplitude of the phase variation, but these models are not
based on any physical model of the instrumental noise, i.e., they are
arbitrary, and they require several additional free parameters that are often
not well constrained by the data or any physical understanding of what these
parameters represent. The overall quality of the decorrelated lightcurve
obtained with these arbitrary and complex models of the the instrumental noise
is also not much better than the best results presented below for the partial
lightcurve. Clearly, a more complete understanding of the instrumental noise
in IRAC for warm mission observations would be a great help for the
interpretation of our data, but in the absence of this we present the results
for the partial lightcurve and make an attempt to quantify the extent to which
instrumental noise introduces systematic errors in our results.

\subsubsection{Optimum decorrelation against position.}
\begin{figure*} 
\begin{center}
\includegraphics[width=0.8 \textwidth]{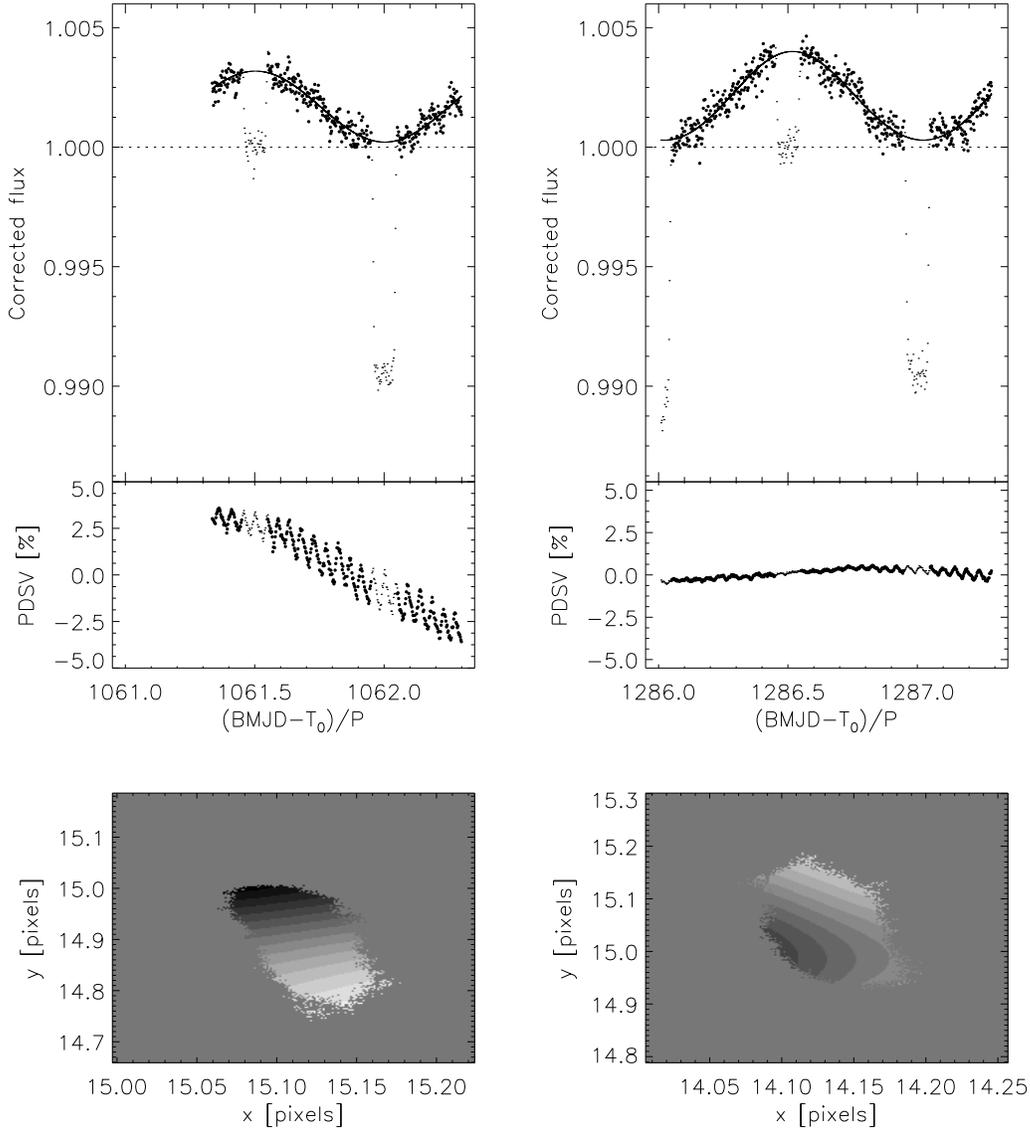} 
\end{center}
\caption{The flux of WASP-18 measured in IRAC channel 1 (left panel) and
channel 2 (right panel) in an aperture of radius 2.25 pixels after correction
for position dependent sensitivity variations (PDSV) for the parameter sets
$(N_{\cos},N_{\sin},N_{x},N_{y},N_{xy})$ = $(2,1,5,10,5)$ and positions
measured with the {\sc cntrd} method. Data are plotted averaged in 200\,s bins
for clarity and the best-fit sinusoidal model is also shown. The mean value in
secondary eclipse is indicated with a dotted line. Note that data in eclipse
(small points) are not included in the fit. The PDSV model is shown as a
function of time in the middle panels and as a function of position as a
grey-scale plot in the lower panels. The grey-scale is linear between $\pm
4$\,per~cent for channel 1 and $1$\,per~cent for channel 2 with positive
values being white. 
\label{decormap}} 
\end{figure*}

We used the model given in equation (1) to fit the data for channel 1 and
channel 2 excluding data within 0.05\,d of mid-transit and mid-eclipse and
also excluding the  first 60\,000 observations (7.3\,hours of data) for
channel 1. We used $N_{\cos} =2 $ and $N_{\sin}=1$ to model the phase
variation of WASP-18. The sine term allows for a phase shift from phase 0.5
for the time of maximum brightness and the first harmonic of the cosine
variations (coefficient $a_2$)  allows for some optimisation of the shape of
this phase variation. To model the PDSV we tried $ N_x = N_{xy} = N_y /2 =
1,2,\dots 6$.  We use $N_y = 2N_x$ because there is a larger range of motion
in the $y$ direction.  We fit these models to the lightcurves for all
combinations of aperture size and position measurement methods.  We used the
Bayesian information criterion (BIC) to identify the combination of
$(N_x,N_y,N_{xy})$ that provides the best compromise between quality of fit
and number of free parameters for a given lightcurve. We calculated the  BIC
using the expression \[ {\rm BIC } = \chi^2 + N_{\rm par}\log_e(N), \] where
$N_{\rm par}$ is the number of free parameters and $N$ is the number of
observations. We used the RMS of the residuals to identify the aperture size
and position measurement method that give the best lightcurves. For both
channels  we find the best results are obtained for $(N_x,N_y,N_{xy}) =
(5,10,5)$ with  positions measured using the {\sc cntrd} method and an
aperture radius of 2.25\,pixels. These models and lightcurves are shown in
Fig.~\ref{decormap} and the parameters of interest are given in
Table~\ref{decortable}. The standard error estimates given in
Table~\ref{decortable}  account for the correlations between parameters
\citep{1992nrfa.book.....P} although the correlation coefficients between the
parameters in this table and all other parameters in the model are small
($<0.3$).

\begin{figure*} 
\begin{center}
\includegraphics[width=0.95 \textwidth]{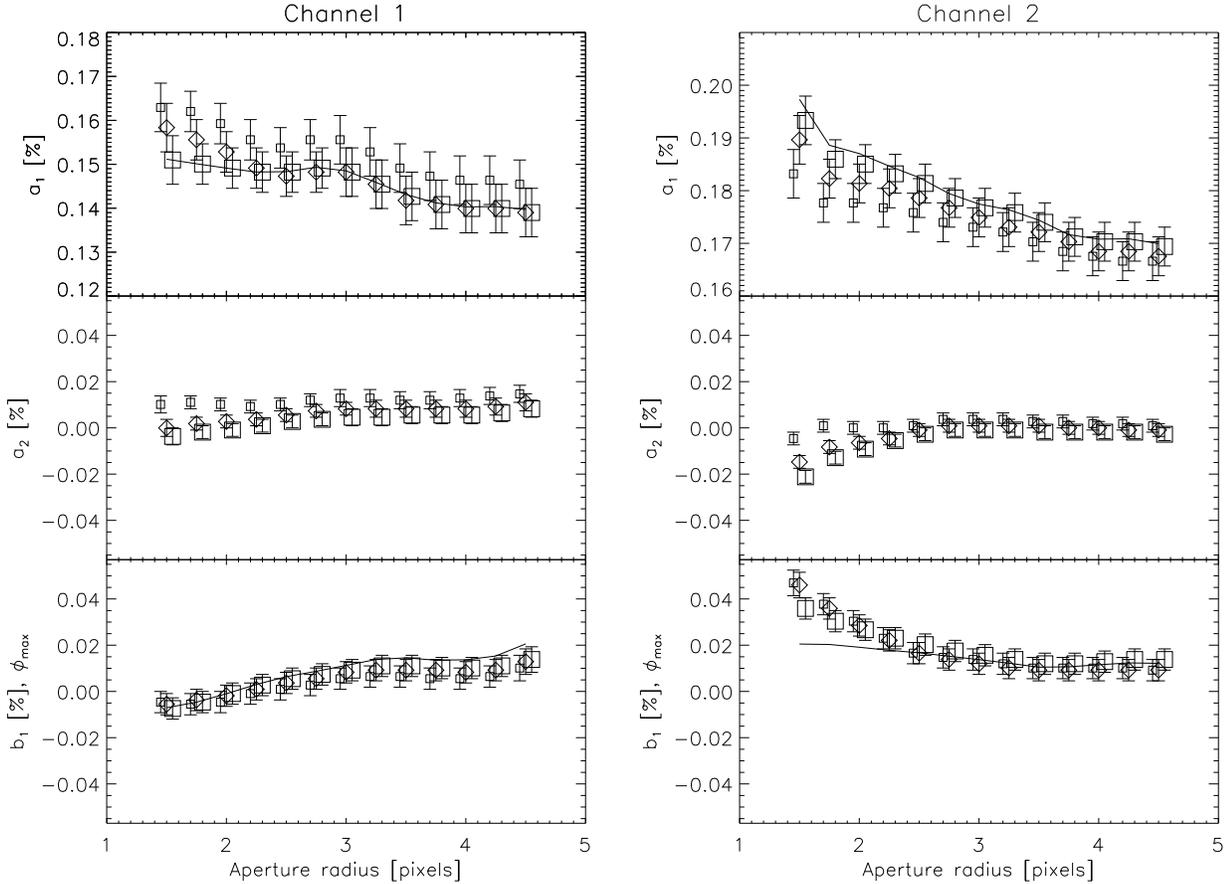} 
\end{center}
\caption{Coefficients of the sinusoidal model for the phase variation in
WASP-18 as a function of aperture radius. The parameter set
$(N_{x},N_{y},N_{xy}) = (5, 10, 5)$ was used for the correction for position
dependent sensitivity. Plotting symbols are as follows: {\sc cntrd} --
squares; {\sc gcntrd} -- diamonds; {\sc gauss2d} -- filled circles. The solid
line in the upper panels shows the semi-amplitude of the phase variation for
the {\sc cntrd} results. The solid line in the lower panel is the
the phase offset from phase 0.5 for time of maximum brightness for the thermal
phase effect derived from the {\sc cntrd} results. Points have been offset
horizontally by $\pm 0.05$ for clarity.
\label{plot215105000}}
\end{figure*}

In Fig.~\ref{plot215105000} we show how the parameters $a_1$, $a_2$ and $b_1$
obtained for $(N_x,N_y,N_{xy}) = (5,10,5)$  vary as a function of aperture
radius and position measurement method for our various lightcurves. Also
plotted in Fig.~\ref{plot215105000} are the amplitude of the phase variation
and the offset from phase 0.5 to the time of maximum brightness in phase
units. There is some dependence on aperture radius for these results, e.g.,
the value of $a_1$ for both channels show a trend towards smaller values with
increasing radius. We also see that there is worse  agreement between the
results for different position measurement methods for smaller apertures as a
result of the increased sensitivity of the pixelation noise to small
differences in the assumed position. For all of the coefficients in both
channels we see that the results vary by about $\pm0.01$\,per\,cent as a
function of aperture radius. We therefore assume that systematic noise limits
the accuracy of these results to $\pm 0.01$\,per\,cent.

Despite the limit of $\pm 0.01$\,per\,cent in the accuracy of these
results, we are able to draw some definite conclusions about the thermal phase
effect in WASP-18. Firstly, the amplitude of the thermal phase effect is very
similar to the depth of the secondary eclipse. This can be seen in
Fig.~\ref{decormap} and by comparing the values for the amplitude in
Table~\ref{decortable} to the eclipse depths given in
Table~\ref{treacletable}. Secondly, the offset between phase 0.5 and the time
of maximum brightness due to the thermal phase effect is consistent with $0$
to within about 0.01 phase units for the channel 1 data and 0.02 phase units
for the channel 2 data. Thirdly, the parameter $a_2$ is also consistent with
the value $0$ so the shape of the thermal phase variation is sinusoidal to
within the limits sets by the systematic noise.

\begin{table}
\caption{Results of linear least squares fits to the phase variation between
eclipses using the model given in equation (1). These results are for
$(N_x,N_y,N_{xy}) = (5,10,5)$ with  positions measured using the
{\sc cntrd} method and an aperture radius of 2.25\,pixels.  $N$ is the number
of points included in the fit and BIC is the Bayesian information criterion as
defined in the text. $A$ is the amplitude of the thermal phase
effect and $\phi_{\rm max}$ is the phase of maximum brightness relative to
phase 0.5. Random  and systematic errors are given for each quantity in that
order.
} 
\label{decortable}
\begin{center}
\begin{tabular}{@{}lrr}
\hline
Parameter & \multicolumn{1}{l}{Channel 1} & \multicolumn{1}{l}{Channel 2}\\
\hline
$a_1$ [\%]    & $ 0.148 \pm 0.005 \pm0.01$ & $ 0.183 \pm 0.004 \pm0.01$  \\
$a_2$ [\%]    & $ 0.003 \pm 0.005 \pm0.01$ & $ 0.023 \pm 0.005 \pm0.01$  \\
$b_1$ [\%]    & $ 0.001 \pm 0.003 \pm0.01$ & $-0.006 \pm 0.003 \pm0.01$  \\
$A$   [\%]    & $ 0.296 \pm 0.009 \pm0.02$ & $ 0.366 \pm 0.007 \pm0.02$ \\
$\phi_{\rm max}$& $ 0.001 \pm 0.003 \pm0.01$ & $-0.010 \pm 0.006 \pm0.02$ \\
$N$             &  133124                   & 179851 \\
$\chi^2$        &    142398.6               & 216717.3 \\
BIC             &    142928.2               & 217259.0 \\
RMS [\%]        &   0.539                   & 0.717\\
\hline
\end{tabular}
\end{center}
\end{table}

\begin{figure*} 
\includegraphics[width=0.9\textwidth]{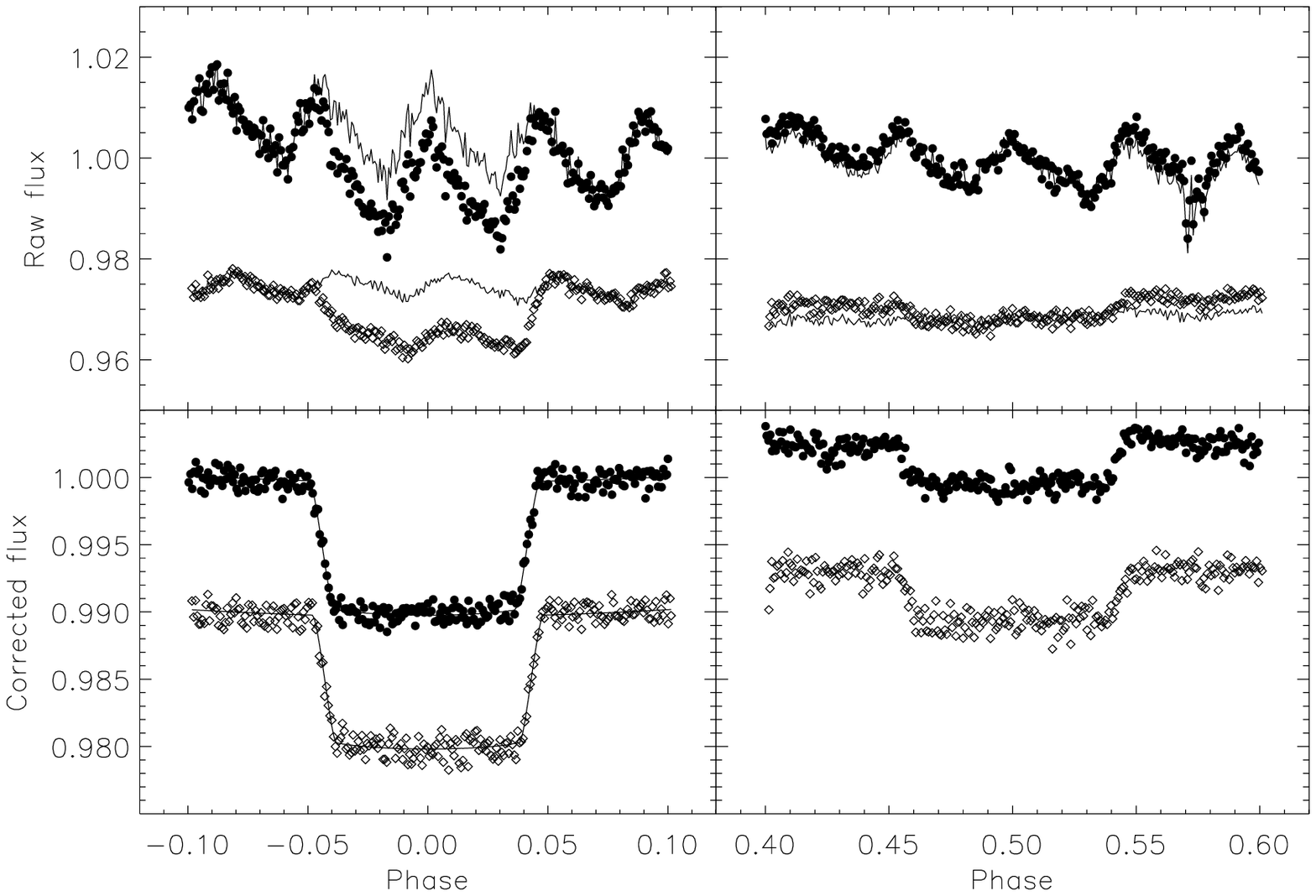} 
\caption{Upper panel: Raw photometry for an aperture radius of 3\,pixels
(filled symbols -- channel 1, open symbols -- channel 2)  together with the
correction for PDSV based on positions measured using the {\sc cntrd} method
(lines). Lower panel: Photometry corrected for PDSV (points) and models fit by
least-squares (lines). The channel 2 data have been vertically offset by 0.03
in the upper panel and 0.01 in the lower panel. In both panels the data and
models are plotted in 60\,s bins. \label{mnmtreaclefit0}} 
\end{figure*}

\subsection{Eclipse model \label{treacle}}
We tried several different methods to model the entire lightcurve for each
channel including both eclipses, the phase variation and PDSV, but were not
able to find any method that gave reliable results. We suspect that there is
some factor other than position on the detector that introduces systematic
noise at the level of  $\sim$0.01\,per\,cent with a timescale of $\sim$day.
This can be seen in Fig.~\ref{decormap}, where there are clear systematic
errors remaining in the lightcurve at this level. These systematic errors are
not removed by increasing the complexity of the model used for the PDSV. It
will be difficult to identify this additional factor given that little
information about the shape of the point spread function can be measured from
the undersampled IRAC images. 

There are many published studies that have used IRAC photometry obtained over
$\sim 5$ hours of observation to successfully model hot Jupiter eclipses, so we
decided to only model the data within 0.1 phase units of the primary and
secondary eclipses. We fit these data simultaneously using a single model to
account for the true flux variations of the system. We then account for
systematic errors in the measured flux independently for the data around each
eclipse. 

We have used the NDE lightcurve model \citep{1972ApJ...174..617N,
1981AJ.....86..102P, 1981psbs.conf..111E} to model the primary and secondary
eclipses in our lightcurves. This model uses biaxial ellipsoids to approximate
to projected area of the star/planet. \cite{2006A&A...450.1231G} has shown
that this model used with an integration ring size of one degree (as we have
done) can be used to model planetary transits with a precision of $\sim
4\times10^{-5}$, which is sufficient for our purposes. From inspection of our
model lightcurves we find that $\sim \frac{1}{4}$ of the model data points
during primary eclipse are affected by numerical noise at this level and that
there is no numerical noise during secondary eclipse. We created a
double-precision version of the NDE model that has negligible numerical noise,
but that runs appreciably slower than the original single-precision code. We
used our double-precision version to verify that the numerical noise  in the
single-precision version has a negligible effect on our results, so all the
results presented here are based on the single-precision version. 

We use the NDE model to calculate $\ell_s(\phi)$ and $\ell_p(\phi)$, the
contribution of the star and planet, respectively, to the total apparent flux
at any given phase, $\phi$, including the effects of tidal distortion and
eclipses. Note that $\ell_s$ and $\ell_p$ include the effects of the eclipses
and transits and the ellipsoidal variation of both star and planet. To model
the variation in magnitude due to the phase effect of the planet we use the
harmonic series \[h(\phi) = a_1\cos(\phi)+b_1\sin(\phi)+a_2\cos(2\phi) -
h_{\rm max},\] where the values of $a_1$, $b_1$ and $a_2$ are taken from the
least-squares fit to the data between transit and eclipse for the same
aperture size and position measurement method and $h_{\rm max}$ is chosen
such that the maximum value (corresponding to minimum flux) of $h(\phi)$ is 0. 

The apparent magnitude of the system is then given by 
\[ m_i =  m_0 - 2.5
\log\left[
\ell_s(\phi_i) + \ell_p(\phi_i) \right] + \ell_p(\phi_i)h(\phi_i)/\ell_{p,\max}, \]
where $\ell_{p,\max}$ is a normalization factor.  Our calculations are done
using magnitudes but we present the results in flux units or as percentages.

In addition, we model the PDSV independently for the data around primary and
secondary eclipse using Legendre polynomial functions of the $x$ and $y$
position plus an optional linear function of time. For each set of lightcurve
model parameters we calculate the optimum values of the PDSV model parameters
using singular value decomposition to fit the residuals from the lightcurve
model.

The parameters of the NDE lightcurve model of relevance to our study are:
$J$, the surface brightness of the planet in units of the central surface
brightness of the star excluding the thermal phase contribution; $r_1 =
R_{\rm star}/a$, the radius of the star in units of the semi-major axis; $r_2
= R_{\rm planet}/a$, the radius of the planet in units of the semi-major axis;
$i$, the inclination, $u_{\star}$, the linear limb-darkening coefficient for
the star; $e\cos(\omega)$ and $e\sin(\omega)$, where $e$ is the orbital
eccentricity and $\omega$ is the longitude of periastron. We fix the mass
ratio of the system at the value $m_{\rm planet}/m_{\rm star} = 0.01$. We did
not use this combination of  parameters directly as free parameters in our
least-squares fitting because there are significant correlations between them.
Instead, we introduce the following parameters which are more directly related
to the observed features of the lightcurve: $\Delta m_{\rm tr}$, $\Delta
m_{\rm oc}$, $W$, $S$.
The parameters of the NDE lightcurve model are then calculated as follows: 
\[k = \frac{r_2}{r_1}  = \sqrt{\frac{\ln(10)}{2.5}\Delta m_{\rm tr}}; \] 
\[r_1 = \frac{\pi}{2\sqrt{k}}\sqrt{W^2(1 - S^2) }; \] \[r_2 = k r_1; \] 
\[b = \sqrt{ \frac{(1-k)^2 - S^2(1+k)^2}{1-S^2} }; \] 
\[i = \cos^{-1}\left(b r_1\right); \] 
\[J = \frac{1-u_{\star}/3}{k^2\left( \frac{2.5}
    {\ln(10)\Delta m_{\rm oc}} - 1\right)}. \]
These parameters are adapted from \cite{2003ApJ...585.1038S} so that, for a
circular orbit, $\Delta m_{\rm tr}$ is the depth of the primary eclipse in
magnitudes, $\Delta m_{\rm oc}$ is magnitude difference between the flux
duration occultation and the minimum of the thermal phase curve; $W$ is the
width of the transit in phase units and $S$ is the duration of the ingress
phase of the transit in units of $W$.  The intermediate variables used here
are $k$, the radius ratio and $b$, the impact parameter. We  also include a
correction to the time of mid-transit compared to the ephemeris of Hellier
et~al., $\Delta T_0$.

We are careful here to define what we mean by the depth of the secondary
eclipse because the variation in flux due to the thermal phase effect on the
timescale of the eclipse is comparable to the precision with which we can
measure the depth from our photometry ($\sim 0.01$\,per\,cent) and the maximum
of the thermal phase effect may not occur at mid-eclipse.  For ease of
calculation, interpretation and comparison with other measurements, we simply
measure the mean flux during occultation (excluding ingress and egress
phases), $f_{\rm in}$,  and the mean flux either side of the eclipse in a
region $\pm$ 0.1 phase units around the time of mid-eclipse, $f_{\rm out}$ and
define the secondary eclipse depth to be $D = \frac{f_{\rm out}-f{\rm
in}}{f_{\rm out}}$. The fluxes are measured from the lightcurve corrected for
PDSV.

\begin{figure} 
\begin{center}
\includegraphics[width=0.49\textwidth]{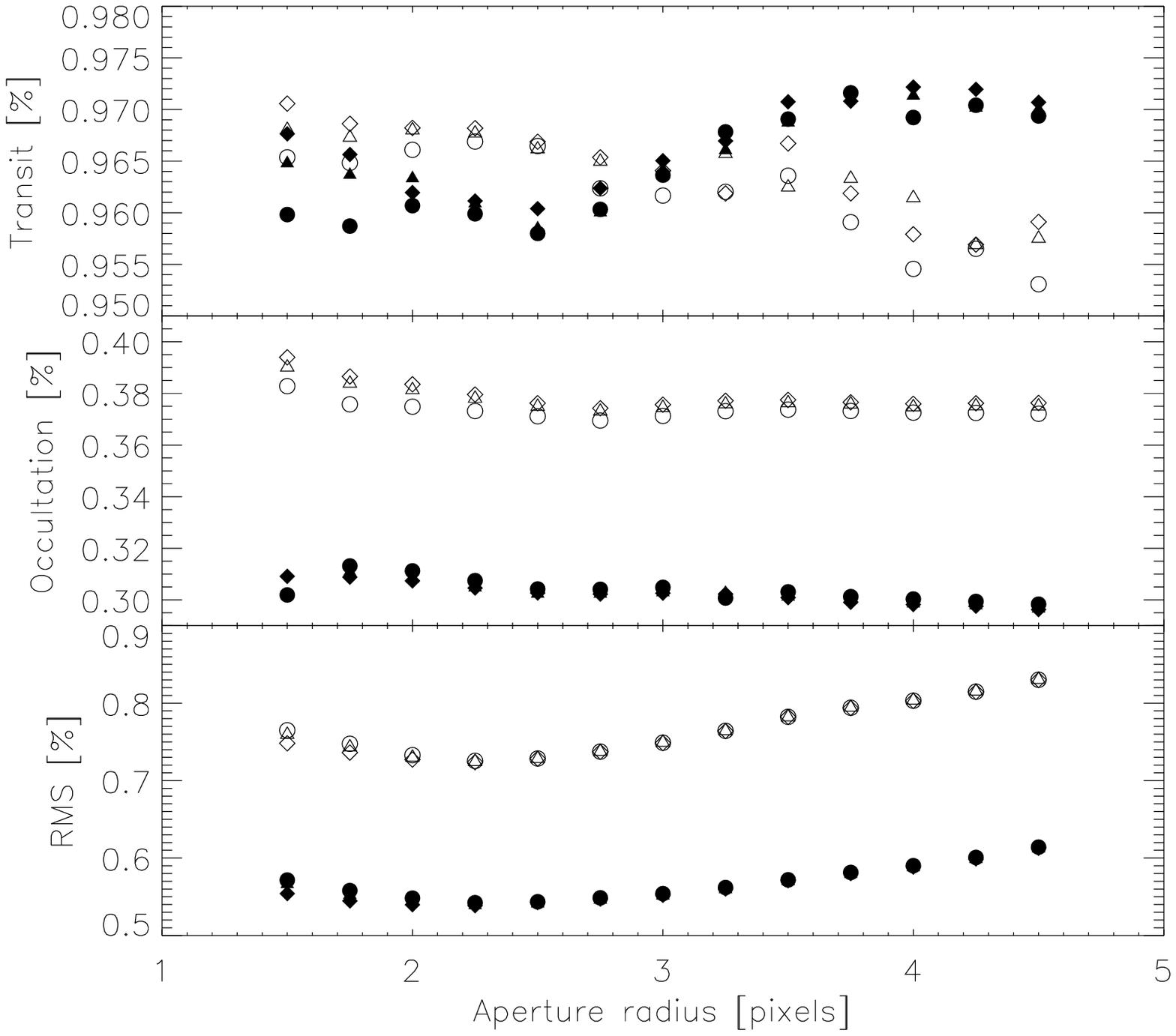} 
\end{center}
\caption{Depths of eclipses for transit and occultation measured by fitting
the eclipses.  Channel 1 data are shown with filled symbols, channel 2 data
with open symbols. Different symbols denote different position measurement
methods.  The RMS of the residuals of the fits are also shown using the same
symbols.
\label{depth0}} 
\end{figure}

There are some second-order effects not accounted for by our model. We do not
account for the brightness distribution on the day-side of the planet, but
this will have a negligible effect on our results given that the thermal phase
effect is not strongly peaked and is symmetrical about phase 0.5, so this
distribution will be approximately uniform and symmetrical. Doppler boosting
is negligible compared to our signal-to-noise ($\la 0.001$\,per\,cent). We
make a small correction to the results for the effects of image
persistence in the channel 2 data by assuming a dilution of the eclipses for
1$\pm$1 \,per\,cent. We have applied a correction to the apparent times of
secondary eclipse for the light travel time across the orbit of $2a/c\approx
20$\,s so that the times and phases  quoted here are the true  time of
mid-occultation relative to the apparent time of mid-transit.  We also apply a
correction to the values of $e\cos(\omega)$ quoted below for this light travel
time. 

We use the simplex algorithm of \citet{simplex} to optimise the least-squares
fit of our model to the lightcurves. The simplex algorithm is a simple way to
optimize a least-squares solution given an initial set of parameters, but it
is not guaranteed to find the global minimum value of $\chi^2$ in the
parameter space. In this case we are able to estimate accurate initial values
for the most important parameters and so any solution will not be very far from 
the global minimum. However, we do find that numerical noise prevents us from
using this algorithm by itself to find the optimum solution. We work around
this problem  by testing many initial starting values. We found that the
solution with the lowest value of $\chi^2$ sometimes has parameters that
are slightly biased when compared to other solutions with similar values of
$\chi^2$ as a result of the numerical noise. We avoid this problem by
taking the median value of each parameter for all solutions with $\chi^2$
within 5 of the minimum as our best estimate of the parameter. 

The results for the depths of the eclipses and the RMS of the residuals for
each data set are shown in Fig.~\ref{depth0}. The depth of the primary eclipse
(transit) can vary slightly with wavelength because the apparent radius of the
planet will be larger at wavelengths where the atmosphere has a large opacity
\citep{2000ApJ...537..916S}. The size of this effect is approximately
$2HR_{\rm planet}/R_{\rm star}^2$ where $H$ is the atmospheric scale height.
In practice, for WASP-18\,b the size of this effect is negligible ($\approx
0.001$\,per\,cent) because the large surface gravity of this massive planet
makes the scale height of the atmosphere ($\sim 40$\,km)  much smaller than
the size of the star. It can be seen from Fig.~\ref{depth0} that the solutions
with the lowest RMS occur for an aperture radius 2 pixels, but the transit
depths for channel 1 and 2 disagree by about 0.01\,per\,cent for these data
sets.

\begin{figure} 
\includegraphics[width=0.48\textwidth]{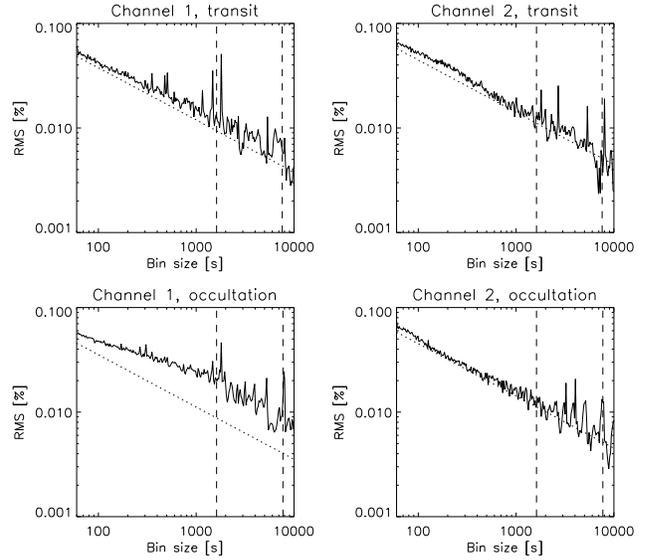} 
\caption{RMS of the residuals after binning as a function of bin size 
(solid line) compared to the predicted photon noise (dotted line). 
The vertical, dashed lined in each panel shows the duration of eclipse and the
duration of ingress/egress.
\label{mnmtreaclenoise0}}
\end{figure}

The transit depths measured in channels 1 and 2 are consistent with each
other for an aperture radius of 3 pixels and lie near the centre of the range
of values obtained. The best fit to the lightcurves for an aperture radius of
3\,pixels using the positions from the {\sc gcntrd} method are shown in
Fig.~\ref{mnmtreaclefit0} and the parameters for the model used are given in
Table~\ref{treacletable}.  It can be seen that there is some residual
correlated noise in the lightcurves after removal of the model for the PDSV.
We quantified this residual correlated noise by calculating the RMS of the
residuals after binning for a range of bin sizes \citep{2006MNRAS.373..231P}.
The results are shown in Fig.~\ref{mnmtreaclenoise0} and compared to the
expectation for pure photon noise. At the timescale of the eclipse
it can be seen that the channel 2 data are only weakly affected
by correlated noise ($\la 0.005$\,per\,cent) but the channel 1 data are
affected correlated noise at a level of 0.005\,--\,0.01\,per\,cent,
particularly for the data covering occultation.

\begin{figure} 
\includegraphics[width=0.5\textwidth]{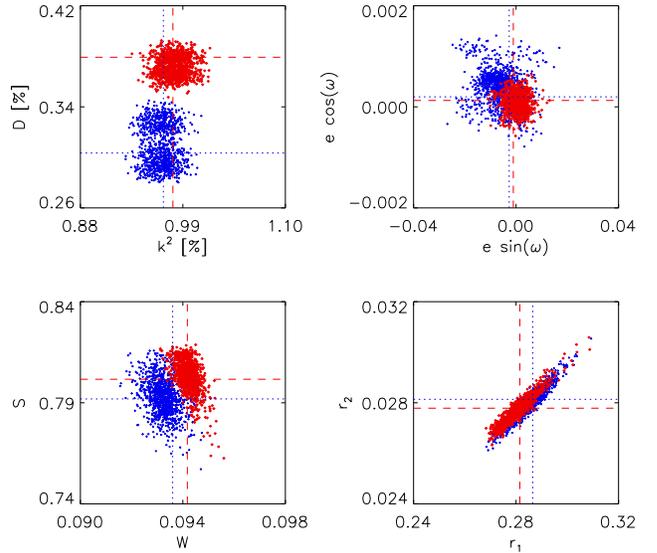} 
\caption{Parameter correlation plots from our residual permutation error
analysis for channel 1 (blue crosses) and channel 2 (red diamonds). Our
adopted values for each channel are indicated using dotted and dashed lines
for channel 1 and 2, respectively. 
\label{mnmtreaclepar}}
\end{figure}

Given that there is significant correlated noise in our data, we decided to
calculate the random error on our model parameters using the ``prayer-bead''
method \citep{ 2006MNRAS.373..231P}. This uses a circular permutation of the
residuals by a random number of steps to create mock data sets. We applied the
circular permutation to the residuals of the primary and secondary eclipses
independently and then used the simplex algorithm to fit models to 1024 mock
data sets. The standard deviation of parameters from the fits is used to
calculate the random errors for the model parameters given in
Table~\ref{treacletable} based on the analysis of the lightcurves for an
aperture radius of 3\,pixels using the positions from the {\sc gcntrd} method.
The random errors quoted include the effect of the uncertainty in correcting
for image persistence in the data. We use the range of values from the
different apertures and position measurement methods to estimate the
systematic errors on each parameter. The distribution of the parameters for
the mock data sets is shown for some parameters of interest in
Fig.~\ref{mnmtreaclepar}. As can be seen from this figure, the eclipse depths
derived from the mock data sets can be biased by up to $\sim 0.005$\,per\,cent
from the actual value. This is an consequence of the correlated noise in the
residuals.  We also used the results from these mock data sets to calculate
the Pearson correlation coefficient, $r$, for all pairs of free parameters
used in the least-squares fit. There is a weak anti-correlation  between $W$
and $S$ ($r\approx -0.5$), and a weak correlation between $W$ and
$e\sin(\omega)$ ($r \approx 0.5$), but the other free parameters are
uncorrelated, as expected. 

\begin{table*}
\caption{Results of least squares fits to the primary and secondary eclipses.
$J^{\prime} = J/(1-u_{\star}/3)$ is the ratio of the integrated surface
brightness of the star and the day-side of the planet.    Other symbols are
defined in the text. Random and systematic errors are given for each parameter
in that order. Parameters that can be derived from the analysis of the optical TRAPPIST
lightcurves are also given in the final column.} 
\label{treacletable}
\begin{center}
\begin{tabular}{@{}lrrr}
\hline
Parameter & \multicolumn{1}{l}{Channel 1} & \multicolumn{1}{l}{Channel 2}&
\multicolumn{1}{l}{TRAPPIST} \\
\hline
$\Delta m_{\rm tr}$ [\%] & $0.969 \pm 0.013  \pm 0.007 $&$0.979 \pm 0.013 \pm 0.009  $ &$ 0.965 \pm 0.056$   \\
$\Delta m_{\rm oc}$ [\%] & $0.015 \pm 0.014  \pm 0.009 $&$0.028 \pm 0.006 \pm 0.017  $ &                     \\
$D$ [\%]                 & $0.304 \pm 0.017  \pm 0.009 $&$0.379 \pm 0.008 \pm0.013   $ &                     \\
$W$                      & $0.0936\pm 0.0005 \pm 0.0003$&$0.0942\pm 0.0003\pm 0.0002 $ &$ 0.0946 \pm 0.0011$ \\
$S$                      & $0.792 \pm 0.009  \pm 0.0007$&$0.802 \pm 0.008 \pm 0.007  $ &                     \\
$u_{\star}$              & $0.06  \pm 0.03   \pm 0.06  $&$0.07  \pm 0.03  \pm 0.04   $ &                     \\
$e\cos(\omega)$          & $0.0002\pm 0.0004 \pm 0.0003$&$0.0001\pm 0.0002\pm 0.0003 $ &                     \\
$e\sin(\omega)$          & $-0.003\pm 0.006  \pm 0.004 $&$-0.001\pm 0.003 \pm 0.002  $ &                     \\
$\Delta T_0$ [s]         & $-109  \pm 8      \pm 0     $&$-108  \pm  8    \pm 0      $ &                     \\
$r_1 $                   & $0.287 \pm 0.006  \pm 0.004 $&$ 0.282\pm  0.005\pm 0.004  $ &$0.291 \pm 0.017$    \\
$r_2$                    & $0.0281\pm 0.0006 \pm 0.0003$&$0.0278\pm 0.0005\pm 0.0005 $ &$0.0286 \pm 0.0026$  \\
$k$                      & $0.0982\pm 0.0007 \pm 0.0004$&$0.0987\pm  0.0006\pm 0.0004$ &$ 0.0983 \pm 0.0030$ \\
$b$                      & $  0.39\pm  0.06  \pm 0.04  $&$ 0.32 \pm  0.06  \pm 0.04  $ &$ 0.41 \pm 0.15$     \\
$i$                      & $ 83.6 \pm   1.0  \pm 0.7   $&$ 84.8 \pm  1.1   \pm 0.8   $ &$ 83  \pm 3$         \\
$J^{\prime}$             & $ 0.33 \pm   0.02 \pm 0.02  $&$ 0.40 \pm  0.01  \pm 0.02  $ &                     \\
Phase of  
~mid-occultation         & $ 0.5003\pm 0.0006\pm 0.0004$ & $0.5002\pm 0.0003\pm0.0005$\\
\noalign{\smallskip}
$N_{\rm transit}$        & 33996 & 35261  & 3649\\
$N_{\rm occultation}$    & 36847 & 34355  &\\
$\chi^2$                 & 74213.9 &  90739.5 & 2768.5\\
RMS [\%]               & 0.55 & 0.75 & 0.40 \\
\hline
\end{tabular}
\end{center}
\end{table*}

\section{Optical variability}
The interpretation of our data would be considerably complicated by any
intrinsic variability of the star WASP-18. 

We have analysed the WASP lightcurves of WASP-18  to determine whether they
show periodic modulation due to the combination of magnetic activity and the
rotation of the star.  The observed value of $V_{\rm rot}\,\sin I = 11
$\,km\,s$^{-1}$ \citep{2009Natur.460.1098H}  together with the stellar radius
imply a rotation period of about  6\,d for WASP-18. We used the sine-wave
fitting method described in \citet{2011PASP..123..547M} to calculate
periodograms over 4096 uniformly spaced frequencies from 0 to 1.5 cycles/day.
The false alarm probability (FAP) for the strongest peak in these periodograms
was calculated using a boot-strap Monte Carlo method also described in
\citet{2011PASP..123..547M}. Variability due to star spots is not expected to
be coherent on long timescales as a consequence of the finite lifetime of
star-spots and differential rotation in the photosphere and so we analysed the
data from each observing season independently.  We removed the transit signal
from the data prior to calculating the periodograms by subtracting a simple
transit model from the lightcurve. In addition to the 2 seasons of data from
\citet{2009Natur.460.1098H} we also analyse 6041 observations obtained during
the 2012 Jun\,--\,Dec observing season. This date range  covers the time of our
Spitzer channel 2 observations.

We did not find any significant periodic signals (FAP$< 0.05$) for WASP-18
apart from frequencies near 1\,cycle/day due to instrumental effects. We
examined the distribution of amplitudes for the most significant frequency in
each Monte Carlo trial and used these results to estimate a 95\,per\,cent upper
confidence limit of 0.1\,per\,cent for the amplitude of any periodic
signal in these WASP lightcurves. \citet{2010MNRAS.409..963B} have shown that
the amplitude of the modulation at IRAC wavelengths due to star spots in
solar-type stars is an order-of-magnitude smaller than at optical wavelengths.
We conclude that any intrinsic variability of WASP-18 due to star spots has a
negligible impact on our analysis.

\section{Eclipse ephemeris}
The analysis above provided two new, precise measurements of the time of
mid-occultation and mid-transit. In addition we have 5 new times of
mid-transit from our TRAPPIST observations. A global analysis of the five
lightcurves was performed with the MCMC software described by Gillon et
al. (2012). In addition to the baseline model and to the transit ephemeris and
shape parameters, the timings of the transits were included as free
parameters, the transit ephemeris being constrained by normal priors based on
the ephemeris presented by \citet{2011ApJ...742...35N}. The details of this
analysis are similar to the ones described in \citet{2012A&A...542A...4G}.
The parameters derived from the least-squares fit to the 5 lightcurves are
shown in Table~\ref{treacletable}. It can be seen that there is very good
agreement between the parameters of the system dervied  from the optical and
infrared lightcurves.

We have also measured a new time
of mid-transit by using a least-squares fit of the NDE lightcurve model to
the 2010 season of WASP data. The time of mid-transit quoted is
close to the mid-point of dates of observation for these data and the
standard error on the time of minimum is calculated using the prayer-bead
method. All these times of mid-eclipse are given in Table~\ref{tmintable}
together with other published times of mid-eclipse.

We used a least squares fit with a single value of the  period and the times
of mid-transit and mid-occultation as free parameters to determine a
following linear ephemeris.
\[{\rm TDB(mid-transit)} =  2455265.5525(1) + 0.9414523(3)\cdot E\]
\[{\rm TDB(mid-occult.)} =  2455266.0234(3) + 0.9414523(3)\cdot E\]

\begin{figure} 
\includegraphics[width=0.45\textwidth]{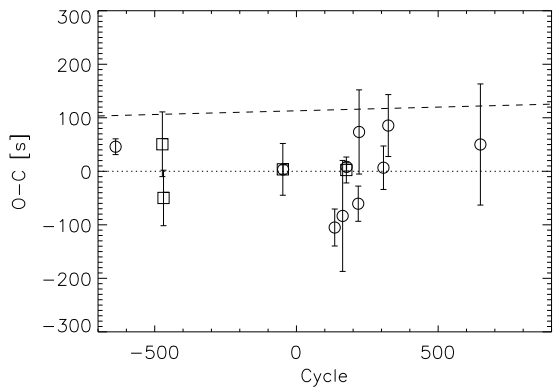} 
\caption{Residuals from our best-fit linear ephemeris for observed times of
mid-transit (circles) and mid-occultation (squares) for WASP-18. The
difference between our ephemeris and the ephemeris of Hellier et~al. is also
shown (dashed line). \label{ephem}}
\end{figure}

The $\chi^2$ value for this fit was 21.8 with 11 degrees-of-freedom so the
standard errors quoted in the final digits here have been scaled by
$\sqrt{21.8/11}$. We also tried a quadratic ephemeris fit to the same data but
found that this did not significantly improve the fit. The residuals from this
$O-C$ diagram for this linear ephemeris is shown in Fig.~\ref{ephem}. It can
be seen that the times of eclipse for WASP-18 have not varied by more than
about 100\,s over 3 years.

\begin{table}
\caption{Apparent Barycentric Dynamical Time (TDB) of mid-transits (tr) and
mid-occultation (oc) for WASP-18.   The cycle number is calculated from our
updated linear ephemeris and  $O-C$ is the residual from this ephemeris. 
Times of mid-occultation have been corrected for light-travel time across the
orbit.
\label{tmintable}}
\begin{center}
\begin{tabular}{@{}rrcrrc}
\hline
\multicolumn{2}{l}{BJD$_{\rm TDB}-2450000$}&
\multicolumn{1}{l}{Type} &
\multicolumn{1}{l}{Cycle} &  
\multicolumn{1}{l}{${\rm O}-{\rm C}$} & 
\multicolumn{1}{l}{Source} \\
\noalign{\smallskip}
\hline
$4664.9061$&$\pm 0.0002$&tr&$-576$&$  0.00013$ &1 \\
$4820.7168$&$\pm 0.0007$&oc&$-411$&$  0.00019$ &2 \\
$4824.4815$&$\pm 0.0006$&oc&$-407$&$ -0.00097$ &2 \\
$5220.8337$&$\pm 0.0006$&oc&$  14$&$ -0.00006$ &a \\
$5221.3042$&$\pm 0.0001$&tr&$  15$&$  0.00017$ &a \\
$5392.6474$&$\pm 0.0004$&tr&$ 197$&$  0.00014$ &3 \\
$5419.0083$&$\pm 0.0012$&tr&$ 225$&$  0.00015$ &b \\
$5431.7191$&$\pm 0.0003$&oc&$ 238$&$ -0.00114$ &a \\
$5432.1897$&$\pm 0.0001$&tr&$ 239$&$ -0.00087$ &a \\
$5470.7885$&$\pm 0.0004$&tr&$ 280$&$ -0.00057$ &c \\
$5473.6144$&$\pm 0.0009$&tr&$ 283$&$  0.00098$ &c \\
$5554.5786$&$\pm 0.0005$&tr&$ 369$&$  0.00026$ &c \\
$5570.5842$&$\pm 0.0006$&tr&$ 386$&$  0.00118$ &c \\
$5876.5559$&$\pm 0.0013$&tr&$ 711$&$  0.00097$ &c \\
\hline
\end{tabular}
\end{center}
1:~\citet{2010A&A...524A..25T}; 2:~\citet{2011ApJ...742...35N}; 3:~{\it
http://var.astro.cz/ETD/}; a:~Spitzer IRAC; b:~WASP; c:~TRAPPIST.
\end{table}

\section{WASP-18 Stellar Parameters}

A total of 21 individual HARPS spectra of WASP-18 were co-added to produce a
single spectrum with a typical S/N of around 200:1. The analysis was performed
using the methods given in \citet{Doyle2012}. The \halpha\ and
\hbeta\ lines were used to give an initial estimate of the effective
temperature (\teff). The surface gravity (\logg) was determined from the
Ca~{\sc i} lines at 6162{\AA} and 6439{\AA} \citep{2010A&A...519A..51B}, along
with the Na~{\sc i} D lines. Additional \teff\ and \logg\ diagnostics were
performed using the Fe lines. An ionisation balance between Fe~{\sc i} and
Fe~{\sc ii} was required, along with a null dependence of the abundance on
either equivalent width or excitation potential \citep{2008A&A...478..487B}.
This null dependence was also required to determine the micro-turbulence
(\mictrb). The parameters obtained from the analysis are listed in
Table~\ref{wasp18-params}. The value of [Fe/H] was determined from equivalent
width measurements of several unblended lines, and additional least squares
fitting of lines was performed when required. The quoted error estimates
include that given by the uncertainties in \teff, \logg, and \mictrb, as well
as the scatter due to measurement and atomic data uncertainties. 

The projected stellar rotation velocity (\vsini) was determined by fitting the
profiles of several unblended Fe~{\sc i} lines. A value for macroturbulence
(\mactrb) of 4.6 $\pm$ 0.3 {\kms} was assumed, based on the calibration by
\cite{2010MNRAS.405.1907B}. An instrumental FWHM of 0.07 $\pm$ 0.01~{\AA} was
determined from the telluric lines around 6300\AA. A best fitting value of
\vsini\ = 10.9 $\pm$ 0.7 ~\kms\ was obtained.

\begin{table}
\caption{Stellar parameters of WASP-18 from our spectroscopic analysis.}
\begin{center}
\begin{tabular}{llr} \hline
Parameter  && \multicolumn{1}{l}{Value} \\ \hline
\teff & [K]  &   6400 $\pm$ 75 \\
\logg  &   &   4.29 $\pm$ 0.10 \\
\mictrb & [\kms] &   1.20 $\pm$ 0.08 \\
\vsini &  [\kms] &   12.1 $\pm$ 0.5  \\
{[Fe/H]}$^{\rm a}$& &   0.10 $\pm$ 0.08 \\
Mass$^{\rm b}$ & [$M_{\sun}$]     &   1.26 $\pm$ 0.09  \\
Radius$^{\rm b}$& [$R_{\sun}$]    &   1.25 $\pm$ 0.15  \\
Sp. Type$^{\rm c}$& &   F6 \\
Distance &[pc] &   130 $\pm$ 20 \\ \hline 
\end{tabular}
\end{center}
\label{wasp18-params}
$^{\rm a}$[Fe/H] is relative to the solar value obtained by
\cite{2009ARA&A..47..481A}.
$^{\rm b}$
Mass and radius estimated using the
\cite{2010A&ARv..18...67T} calibration.
$^{\rm c}$ Spectral Type estimated from \teff\
using the table in \cite{2008oasp.book.....G}. 
\end{table}

The rotation rate ($P = 5.8 \pm 0.8$~d) implied by the {\vsini} gives a
gyrochronological age of $\sim 1.1^{+4.7}_{0.6}$~Gyr using the
\citet{2007ApJ...669.1167B} relation. The value of \teff\ derived from our
spectroscopic analysis agrees well with the value $6455\pm70$\,K derived by
\citet{2011MNRAS.418.1039M} from optical and near-infrared photometry using
the infrared flux method. The distance derived here assuming that WASP-18 is a
main-sequence star and quoted in Table~\ref{wasp18-params} is consistent with
the value $100\pm10$\,pc derived from the Hipparcos parallax
\citep{2007A&A...474..653V}.

\section{Physical parameters}

\begin{table} 
\caption{\label{tab:wasp18:model} Derived physical properties of the WASP-18
system. Parameter values are shown with random and, where appropriate,
systematic errors, respectively. }
\begin{center}
\begin{tabular}{l r@{\,$\pm$\,}c@{\,$\pm$\,}l }
\multicolumn{1}{l}{Parameter} & \multicolumn{1}{c}{Value} \\
\hline
$M_{\rm A}$    (\Msun) & 1.295    & 0.052    & 0.027       \\
$R_{\rm A}$    (\Rsun) & 1.255    & 0.027    & 0.009       \\
$\log g_{\rm A}$ (cgs) & 4.353    & 0.017    & 0.003       \\
$\rho_{\rm A}$ (\psun) & \mcc{$0.655 \pm 0.035$}           \\[2pt]
$M_{\rm b}$    (\Mjup) & 10.52    &  0.28    &  0.15       \\
$R_{\rm b}$    (\Rjup) & 1.204    & 0.027    & 0.008       \\
$g_{\rm b}$     (\mss) & \mcc{$179.9 \pm   6.4$}           \\
$\rho_{\rm b}$ (\pjup) & 5.64     & 0.31     & 0.04        \\[2pt]
\Teq\              (K) & \mcc{$2411 \pm   35$}             \\
$a$               (AU) & 0.02055  & 0.00028  & 0.00014     \\
Age              (Gyr) & \ermcc{0.4}{0.8}{0.9}{0.5}{0.3}   \\
\hline \end{tabular} 
\end{center}
\end{table}

Our new photometric and spectroscopic results allow for an improved
determination of the physical properties of the WASP-18 system. We performed
this analysis following the method of \citet{2009MNRAS.394..272S}, which
requires as its input parameters measured from the lightcurves and spectra,
plus tabulated predictions of theoretical models. From the lightcurves we
adopted $r_1 = 0.284 \pm 0.005$, $r_2 = 0.0280 \pm 0.0005$ and $i = 84^\circ
\pm 1^\circ$. The stellar \teff\ and [Fe/H] were taken from the
spectroscopic determination in the previous section, and the star's velocity
amplitude was taken to be $K_1 = 1816.7 \pm 1.9$\,m\,s$^{-1}$
\citep{2010A&A...524A..25T}.

An initial value of the velocity amplitude of the planet, $K_2$, was used to
calculate the physical properties of the system with  the physical constants
listed by \citet {2011MNRAS.417.2166S}. The mass and [Fe/H] value  of the star
were then used to obtain the expected \teff\ and radius, by interpolation
within one set of tabulated predictions from stellar theory. $K_2$ was refined
in order to find the best agreement between the observed and expected \teff,
and the measured $r_1$ and expected $\frac{R_1}{a}$. This was performed for
ages ranging from the zero-age main sequence to when the star was
significantly evolved ($\logg < 3.5$), in steps of 0.01\,Gyr. The overall best
fit was found, yielding estimates of the system parameters and also the
stellar age.

This procedure was performed separately using five different sets of stellar
theoretical models \cite[see][for details]{2010MNRAS.408.1689S} plus a
calibration of stellar properties based on well-studied eclipsing binary star
systems \citep{2010A&A...516A..33E}, with calibration coefficients from
\citet{2011MNRAS.417.2166S}. The results are given in
Table~\ref{tab:wasp18:model}, where we quote the mean value for each
parameter, the random error and an estimate of the systematic error  from the
range of values derived from the  different stellar models, where appropriate.
It can be seen from Table~\ref{tab:wasp18:model} that the results from
different models are consistent to within the random errors on each parameter.

In comparison to previous work, we have derived more precise radii, surface
gravities and densities, for both components. We constrain the age of the star
to be less than 1.7\,Gyr, consistent with the gyrochronological age derived
above.

\section{Possibility of contamination by a companion star}

We have estimated the probability that our Spitzer photometry of WASP-18 is
contaminated by the ``third-light'' from a companion star. It is not possible
to detect a modest amount of third-light contamination directly from the
lightcurve itself because its only effect is to reduce the depths of the
eclipses. It would be possible to find a good fit to a lightcurve affected by
third-light contamination, but the parameters of the lightcurve model would be
biased, e.g., $k$ would be too small.

Our calculation is based on the upper limit from our AO observations of 4.0
magnitudes for the brightness of any companion between 0.2\,--\,2\,arcseconds
from WASP-18 and the upper limit of 43\,m\,s$^{-1}$\,y$^{-1}$ over a baseline
of 500\, days to the variation in the mean radial velocity of WASP-18 from
\citet{2010A&A...524A..25T}. We assume that the probability distribution for
the mass, eccentricity and period of the hypothesised companion is the same as
the distributions for companions to solar-type stars from
\citet{2010ApJS..190....1R}. We approximated the distribution of companion
masses using a uniform distribution from 0.2\,\Msolar\ to 0.8\,\Msolar\ and
used a uniform eccentricity distribution from 0 to 1. We then created a set of
65536 simulated binary stars with randomly selected periods, masses and
eccentricities according to these distributions and randomly orientated
orbits. We found that of these simulated binary stars, approximately
55\,per\,cent would have been resolved by our AO imaging at the distance of
WASP-18; 20\, per\,cent would have orbital periods less than 500\, days and a
semi-amplitude of 43\,m\,s$^{-1}$ or more and 25\,per\,cent would show a
change in radial velocity of 43\,m\,s$^{-1}$ or more over 500 days. This
leaves only 5\,per\,cent of hypothesised binaries as not detectable given our
AO imaging and the published radial velocity data. The probability that
WASP-18 has a stellar companion is further reduced because the overall binary
fraction observed for planet hosting stars is approximately 25\,per\,cent
\citep{2010ApJS..190....1R}. An M-dwarf at the same distance as WASP-18 just
below out detection limit of 4.0\,magnitudes in the K-band at 0.2\,arcsec
would contribute no more than 5\,per\,cent of the light at 4.5\,$\mu$m. The
more stringent limit of 6.0 magnitudes in the K-band that applies for
separations of 0.5\,--\,2.0\,arcsec corresponds to an M-dwarf that contributes
no more than 1\,per\,cent at 4.5\,$\mu$m. Of the simulated binary stars
approximately 45\,per\,cent would be detected at this resolution.

 In conclusion, our AO imaging and the published radial velocity data show
that it is unlikely that WASP-18 has a stellar companion that significantly
contaminates our Spitzer photometry.

\section{Discussion}

 The values of $D$ in Table~\ref{treacletable} are in very good agreement the
values $0.31\pm 0.02$ at 3.6\,$\mu$m and 0.38$\pm0.02$\,per\,cent at
4.5\,$\mu$m measured independently by \citet{2011ApJ...742...35N}. Their
analysis of the secondary eclipse depths in 4 IRAC passbands suggests that the
day-side atmosphere of WASP-18 is likely to feature a temperature inversion.
For zero albedo and zero redistribution of heat to the night side of the
planet the integrated brightness temperature for the day-side is
$T_{\varepsilon=0} = \left(\frac{2}{3}\right)^{1/4}T_0 = 3110 \pm 35$\,K
\citep{2011ApJ...729...54C}. For black-body emission this implies eclipse
depths of $0.329\pm0.005$ at 3.6\,$\mu$m and $0.379 \pm 0.011$\,per\,cent at
4.5\,$\mu$m, both in good agreement with the observed values. 
Zero-redistribution of heat within the atmosphere is also consistent with our
observation that the peak of the thermal phase curve occurs close to the time
of mid-occultation. Little can be said about the chemical composition of the
day-side atmosphere at this stage because no  strong molecular absorption or
emission features have been detected  from these secondary eclipse depth
measurements.

 The amplitude of the thermal phase curve we have measured and the lack of a
significant offset between the maximum in this curve and the time of
mid-eclipse are both consistent with the conclusion based on the secondary
eclipse depths that the albedo and recirculation efficiency for WASP-18 are
both very low. This is consistent with the hypothesis that very hot Jupiters
have weak recirculation based mainly on secondary eclipse depth measurements
alone \citep{2011ApJ...729...54C}. The good agreement between the
recirculation efficiency inferred from the eclipse depths and measured from
the thermal phase curve for WASP-18 strengthens this conclusion.

 The stellar limb darkening at infrared wavelengths is lower than at
optical wavelengths and so the transit produces a more ``box-shaped'' eclipse. 
This, combined with the precise photometry that is possible with Spitzer IRAC
data, results in more precise estimates for parameters such as $R_{\rm
star}/a$, $R_{\rm planet}/a$ and $k = R_{\rm planet}/R_{\rm star}$. Our
results for these parameters agree well with the results of
\citet{2009ApJ...707..167S}. The agreement with the results of
\citet{2010A&A...524A..25T} is less good mainly because they find a larger
stellar radius than our study ($R_{\rm star}/a = 0.313\pm0.010$).
The values of  $e\cos(\omega)$ and $e\sin(\omega)$ derived from our analysis
agree well with those of \citeauthor{2010A&A...524A..25T}, but the value of
$e\sin(\omega) = 0.0085\pm0.0009$ they derive from their high quality radial
velocity data is much more precise than ours and points to a small but
significantly non-zero eccentricity. \citet{2012MNRAS.422.1761A} have argued
that the small value of this apparent eccentricity combined with longitude of
periastron very close to $\omega=90^{\circ}$ is exactly the signal expected
due to surface flows induced by tides on the planet. Their conclusion that the
orbital eccentricity of WASP-18\,b is less than 0.009 is consistent with the
results of our analysis, although we are not able to confirm whether
$e\ll0.009$ as they suggest.

 The measurement of the thermal phase effect for hot Jupiters using a
continuous set of observations over an orbital cycle with  Warm
Spitzer is not a well-established technique, so it is useful to compare our
experience of observing WASP-18 with the results using a
similar observing strategy for WASP-12 obtained by \citet{2012ApJ...747...82C}
and for HD\,189733 by \citet{2012ApJ...754...22K}. We find that systematic
errors of unknown origin limit the accuracy with which we can measure the
amplitude of signals with time-scale comparable to the orbital period to about
$\pm 0.01$\,per\,cent.  The main difficulty that
\citeauthor{2012ApJ...747...82C} report in their WASP-12 analysis is a signal
on twice the orbital frequency in the 4.5\,$\mu$m data that they tentatively
attribute to the ellipsoidal modulation of WASP-12\,b. However, as they make
clear, this signal is not seen in their 3.6\,$\mu$m data and may be due to
``uncorrected systematic noise''. The amplitude of this ``$\cos(2\phi)$''
signal in their 4.5\,$\mu$m data is about $\pm 0.1$\,per\,cent, ten times
larger than the level of systematic noise we find on these timescales. We do
not see any signal for the $\cos(2\phi)$ harmonic in our 4.5\,$\mu$m data
greater than about 0.02\,per\,cent.  \citeauthor{2012ApJ...754...22K} use the
wavelet-based method of \citet{2009ApJ...704...51C} to account for systematic
noise in their full-orbit lightcurves of HD\,189733 by assuming that this
noise has a power spectral density varying as 1/frequency. They find that the
systematic noise contributes 0.0162\,per\,cent  of the total scatter in their
channel 1 data -- comparable to the level seen in our data -- but only
0.0017\,per\,cent in channel 2 -- much less than we see in our data. A
systematic application of the wavelet-based method to archival Spitzer data
may be a useful way to better understand the systematic noise sources in this
instrument and perhaps identify observing strategies that can reduce
systematic noise levels.

\section{Conclusions}

 The amplitude, shape and phase of the thermal phase effect we have measured
from our Warm Spitzer lightcurves of WASP-18  are consistent with a sinusoidal
variation with the same amplitude as and symmetric about the secondary
eclipse, to within an accuracy $\approx$ 0.01\,per\,cent set by some
unknown source of systematic error. One contribution to this systematic error
is likely to be the image persistence we observe from the offset images
obtained immediately after our observations of WASP-18. This leads to the
conclusion that WASP-18\,b has a low albedo and that heat transport to the
night-side of the planet is inefficient. This is the same conclusion reached
by \citet{2011ApJ...742...35N} based on the eclipse depths at 3.6\,$\mu$m,
4.5\,$\mu$m, 5.8\,$\mu$m and 8.0\,$\mu$m. The eclipse depths we measure at
3.6\,$\mu$m and 4.5\,$\mu$m are consistent with the previous measurements.

\section*{Acknowledgements} 
This work is based on observations made with the Spitzer Space Telescope,
which is operated by the Jet Propulsion Laboratory, California Institute of
Technology under a contract with NASA. Support for this work was provided by
NASA through an award issued by JPL/Caltech. PM would like to thank Felipe
Menanteau for providing his proprietary data to us for the analysis of the
image persistence artifacts. We thank Bryce Croll and Heather Knutson for
enabling the AO observations of WASP-18 to be obtained and included in this
manuscript.
 
\label{lastpage} \bibliographystyle{mn2e}
\bibliography{wasp}

\end{document}